\documentclass[aps,twocolumn,nofootinbib]{revtex4}

\usepackage{graphicx}
\usepackage{dcolumn}
\usepackage{bm}

\usepackage{amssymb,amsfonts}
\usepackage{epsfig}
\usepackage{epstopdf}
\usepackage[all]{xy}
\usepackage{amsthm}
\usepackage{dcolumn}
\usepackage{hyperref}
\usepackage{url}
\usepackage{dsfont}
\usepackage{slashed}
\usepackage{mathrsfs}

\newcommand{\be}{\begin{equation}}
\newcommand{\ee}{\end{equation}}
\newcommand{\bea}{\begin{eqnarray}}
\newcommand{\nn}{\nonumber}
\newcommand{\eea}{\end{eqnarray}}

\def\inbar{\,\vrule height1.5ex width.4pt depth0pt}
\def\IR{\relax{\rm I\kern-.18em R}}
\def\IC{\relax\hbox{$\inbar\kern-.3em{\rm C}$}}

\begin{document}

\title{``Massive" Rarita-Schwinger field in de Sitter space}

\author{Hamed Pejhan$^{1}$\footnote{pejhan@zjut.edu.cn}}

\author{Mohammad Enayati$^2$}

\author{Jean-Pierre Gazeau$^{3}$\footnote{gazeau@apc.in2p3.fr}}

\author{Anzhong Wang$^{1,4}$\footnote{Anzhong\_Wang@baylor.edu}}

\affiliation{$^1$Institute for Theoretical Physics and Cosmology, Zhejiang University of Technology, Hangzhou 310032, China}

\affiliation{$^2$Department of Physics, Razi University, Kermanshah 6741414971, Iran}

\affiliation{$^3$APC, Univ Paris Diderot, Sorbonne Paris Cit$\mbox{\'{e}}$ Paris 75205, France}

\affiliation{$^4$GCAP-CASPER, Physics Department, Baylor University, Waco, TX 76798-7316, USA}

\begin{abstract}
We present a covariant quantization of the ``massive" spin-${\frac{3}{2}}$ Rarita-Schwinger field in de Sitter (dS) spacetime. The dS group representation theory and its Wigner interpretation combined with the Wightman-G$\mbox{\"{a}}$rding axiomatic and analyticity requirements in the complexified pseudo-Riemanian manifold constitute the basis of the quantization scheme, while the whole procedure is carried out in terms of coordinate-independent dS plane waves. We make explicit the correspondence between unitary irreducible representations (UIRs) of the dS group and the field theory in dS spacetime: by ``massive" is meant a field that carries a particular principal series representation of the dS group. We drive the plane-wave representation of the dS massive Rarita-Schwinger field in a manifestly dS-invariant manner. We show that it exactly reduces to its Minkowskian counterpart when the curvature tends to zero as far as the analyticity domain conveniently chosen. We then present the Wightman two-point function fulfilling the minimal requirements of local anticommutativity, covariance, and normal analyticity. The Hilbert space structure and the unsmeared field operator are also defined. The analyticity properties of the waves and the two-point function that we discuss in this paper allow for a detailed study of the Hilbert space of the theory, and give rise to the thermal physical interpretation.
\end{abstract}
\maketitle

\section{Introduction}
Quantum field theory (QFT) in dS spacetime has been a subject of growing interest during the last four decades. In the 1970s, the attention was mainly because of the large isometry group of this spacetime: the dS solution to the cosmological Einstein equations (with positive cosmological constant) has the same degree of symmetry as the Minkowski solution and it can be viewed as a one-parameter deformation of the latter, involving a fundamental length (the dS radius) $R$. In this sense, dS spacetime has been always respected during the past 40 years as an excellent laboratory offering a guideline to perform the otherwise difficult task of quantizing fields in more elaborate gravitational backgrounds. Moreover, the radius $R$ has been regarded as providing a (dS-covariant) infrared cutoff for Minkowskian QFT's, whose removal regenerates automatically Poincar$\mbox{\'{e}}$ covariance.

In the 1980s, there was a great revival of interest for models of QFT in dS spacetime, when it turned out that the dS metric plays a critical role in the inflationary cosmological scenario. Based on the latter, our Universe underwent a dS phase in the very early epochs of its life \cite{Linde}. A possible explanation of phenomena occurring in the very early Universe then relies on an interplay between spacetime curvature and thermodynamics and an outstanding role is played by the mechanisms of symmetry breaking and restoration in a dS QFT.

In the late 1990s, two other remarkable events led to renewed interest in QFT in dS and asymptotically dS spacetimes. On one hand, astrophysical data coming from type Ia supernovae \cite{Perlmutter} indicated that the cosmic expansion is accelerating and pointed towards the existence of a small but non-vanishing positive cosmological constant. This means that our Universe, beside the very early epochs of its life, might currently be in a dS phase which approaches to a pure dS spacetime. [$R=cH^{-1}$ in which $H$ is the Hubble constant fixing the rate of expansion of the spatial sections.] On the other hand, success of AdS/CFT correspondence \cite{Maldacena} has led to intense study to obtain an analogous correspondence in the dS case (the dS/CFT correspondence) \cite{Hull,Witten,Strominger}.

All of these developments plead in favor of setting up a model of QFT in dS spacetime with the same level of completeness and rigor as its Minkowskian counterpart. In this regard, we refer in particular to a promising formulation of such a theory and its subsequent thermic interpretation that was originally put forward for the ``massive" scalar fields in dS spacetime in the 1990s \cite{Bros,BrosPRL,BrosComm},\footnote{The quantization of a scalar field on dS space was first described by Chernikov \cite{Chernikov1968} and Tagirov \cite{Tagirov73} and has also been studied by Schomblond \cite{Schom68, Schom76}, Mottola \cite{Mottola84} and Allen \cite{Allen85}.} and during the last two decades, it has been subject to scrutiny in a number of works to make explicit the extra algebraic structure inherent to other dS \emph{elementary systems} (see, for instance, \cite{de Bievre6230,masive/vect,Gazeau1415,Diracfields,massive/2,massless/2,Garidi/2005,massless/22,dSGravityII,dSGravityI,BambaI,massless/vect}).

Technically, this model of dS QFT enjoys a robust group theoretical content. As a matter of fact, by ``massive" field in this model is meant an object which has a non-ambiguous massive limit as spacetime becomes flat. Group representation theory and its Wigner interpretation according to elementary system indeed allow for control of this limiting procedure through contraction of group representation \cite{Mickelsson,Dooley,GHR,barut,What}. Accordingly, the dS group ``massive" representations would be those of the principal series of the dS representations which contract to the Poincar$\mbox{\'{e}}$ group massive ones, and consequently, the dS massive fields would be those which transform under the principal series representations. The situation for the dS ``massless" fields, however, is different. They are specified by reference to conformal invariance and propagation on the dS light-cone. Hence, it is natural to call ``massless" representations of the dS group $SO_0(1,4)$ as the ones naturally extended to the conformal group $SO_0(2,4)$ \cite{AFFS}. Based on this criterium, one finds that the dS massless scalar field transforms under a specific representation of the complementary series of the dS group, while the massless spinorial cases transform under the dS representations lying at the lower end of the discrete series. On this basis, having well-defined massive and massless dS fields in the context of group representation theory, one can encounter the corresponding \emph{covariant} QFTs along the lines suggested by Wightman and G$\mbox{\"{a}}$rding in their seminal paper \cite{WG}. Here, we must underline that this formulation of dS QFT based on the dS group representation theory and its Wigner interpretation features a remarkable advantage that it would be \emph{not} coordinate dependent.

This quantization scheme is eventually supplemented by analyticity properties offered by the complexified pseudo-Riemanian manifold, in which the dS manifold is embedded. These properties lead to a new \emph{plane-wave}\footnote{The dS plane waves, independent of the choice of the coordinate system, simply allow one to manage the dS group representations and the Fourier transform in the flat limit.} representation of the two-point functions, paving the way to a general momentum space analysis for dS QFT. The analyticity properties have appeared to be crucial from both computational and conceptual points of view. As a matter of fact, all the QFT's considered have a thermal interpretation in view of the existence of a temporal curvature of a very specific nature (as in the Unruh effect \cite{Sewell,Unruh} and in the black-hole evaporation \cite{Sewell,Hawking}). In this sense, the existence of complex hyperbolic trajectories on which maximal analyticity properties hold consistently with locality is a geometric criterion for QFT \emph{vacua} in which thermal effects are produced, the temperature being proportional to the curvature of these trajectories.

In the present paper, motivated by this solid framework, we proceed to the quantization of the massive spin-$\frac{3}{2}$ Rarita-Schwinger field in dS spacetime. This theory is interesting in itself since it obviously is a step that should be taken in order to formulate interacting QFTs. On the other hand, higher spin elementary particles (spin $\geq {\frac{3}{2}}$), such as the gravitino (if it exists, it is a fermion of spin ${\frac{3}{2}}$), play an important role in supersymmetry,\footnote{In fact, it is possible to construct a theory of gravity possessing local supersymmetry only when the massless gravitino exists. The massive gravitino arises in theories where supersymmetry is broken, and the gravitino gets mass by the super Higgs mechanism \cite{Basu}.} which itself represents a fundamental building block of many modern unification schemes. Furthermore, knowledge of the dS massive fields enables us to display the structure of the field theory that appears at the massless limit. Massless field theories in dS spacetime possess some peculiarities that have no analog in flat case (see for instance \cite{massless/vect,Flato1978,Hidden}). Having an expression for the dS massive fields permits one to determine the indecomposable representation describing the dS massless fields.

To achieve our goal, the rest of the paper is organized as follows. We begin our discussion in Section II by presenting the dS machinery, space, group and representation. We use the dS ambient space formalism which constitutes a coordinate-independent approach and makes apparent the group theoretical content of the model. In Section III, admitting the spinor-vector representation of the dS massive spin-$\frac{3}{2}$ field, the first-order dS Rarita-Schwinger field equation is given. In Section IV, we drive the general solutions to the field equation in terms of dS spinor-vector plane waves. The latter are singular on lower-dimensional subsets in dS spacetime: therefore, they are only locally defined. To circumvent this difficulty, the dS plane waves must be considered as distributions which are boundary values of analytic continuations of the solutions to tubular domains in the complexified dS space. On this basis, we particularly point out that the driven dS massive Rarita-Schwinger field reduces to the usual one in the flat limit of the theory: no negative energy appears in the limit process. In Section V, we focus on the corresponding Wightman two-point function and its analyticity properties that imply a KMS thermal interpretation. We drive the Wightman two-point function $S^{(\frac{3}{2})}_{\alpha\alpha^\prime}(x,x^\prime)$ in terms of the dS plane waves, while it enjoys the conditions of: \emph{i}) positiveness, \emph{ii}) local anticommutativity, \emph{iii}) covariance, \emph{iv}) transversality, \emph{v}) divergencelessness, \emph{vi}) tracelessness, and \emph{vii}) normal analyticity. The normal analyticity allows us to define the two-point function $S^{(\frac{3}{2})}_{\alpha\alpha^\prime}(x,x^\prime)$. Then, we make the Hilbert space structure explicit. As a matter of fact, the explicit knowledge of the two-point function with the above-mentioned properties allows us to construct an acceptable quantum theory of the dS Rarita-Schwinger field. Finally, we summarize our results in Section VI.

\section{Presentation of the dS machinery}

\subsection{$1+3$-dS geometry and kinematics}
Geometrically, dS spacetime can be viewed as (the covering space of) the one-sheeted four-dimensional hyperboloid ${\cal M}_H$ embedded in a five-dimensional Minkowski space $\mathbb{R}^5$,
\begin{eqnarray}
{\cal M}_H &=& \{ x\in{\mathbb{R}^5}:\;x^2= x\cdot x =\eta_{\alpha\beta}x^\alpha x^\beta=-H^{-2} \},\nonumber\\
\alpha, \beta &=& 0,1,2,3,4, \;\;\; \eta_{\alpha\beta}=\mbox{diag}(1,-1,-1,-1,-1), \nonumber
\end{eqnarray}
with the notations $x^\alpha \equiv (x^0,\vec{x},x^4)$. The dS line element is obtained concretely by inducing the natural metric on the dS hyperboloid,
\begin{eqnarray}
ds^2= \eta_{\alpha\beta}dx^{\alpha}dx^{\beta}|_{x^2=-H^{-2}} = g^{dS}_{\mu\nu} dX^{\mu}dX^{\nu},
\end{eqnarray}
where the coordinates $X^\mu$'s, $\mu = 0,1,2,3$, are the four local spacetime coordinates for the dS hyperboloid.

A global causal ordering (induced from that of the ambient spacetime $\mathbb{R}^5$) exists on the dS manifold ${\cal M}_H$: technically, let
\begin{eqnarray}\label{V^+}
\overline{V^+} = \{x\in\mathbb{R}^5:\; x\cdot x \geq 0,\; \mbox{sgn}\;x^0 = + \},
\end{eqnarray}
be the future cone in $\mathbb{R}^5$, then considering two events $x,x^\prime$ in ${\cal M}_H$, $x$ would be future connected to $x^\prime$, i.e., $x\geq x^\prime$, if and only if $x-x^\prime \in \overline{V^+}$. The closed causal future (respectively, past) cone of a given point $x\in {\cal M}_H$ is denoted by the set $\Upsilon ^+(x)$ (respectively, $\Upsilon ^-(x)$),
\begin{eqnarray}
\Upsilon ^+(x) &=& \{x^\prime\in {\cal M}_H:\; x^\prime\geq x\},\nonumber\\
\Upsilon ^-(x) &=& \{x^\prime\in {\cal M}_H:\; x^\prime\leq x\}.
\end{eqnarray}
We say two points $x,x^\prime \in {\cal M}_H$ are in ``acausal relation" or ``space-like separated" if $x^\prime\notin \Upsilon^+(x)\cup \Upsilon^-(x)$, i.e., if $x\cdot x^\prime > -H^{-2}$.

The relativity group of the dS spacetime, as the Lorentz group of the ambient Minkowski space, is denoted by $SO_0(1,4)$. This group leaves invariant the quadratic form $x\cdot x = x_0^2 - x_1^2 - ... -x_4^2$ and each of the sheets of the cone ${\cal C}={\cal C}^+\cup {\cal C}^-$, where
\begin{eqnarray}
{\cal C}^\pm=\{x\in \mathbb{R}^5:\;x^2=0,\; \mbox{sgn}\;x^0 =\pm \}.
\end{eqnarray}
The dS group acts transitively on ${\cal M}_H$; therefore, one can distinguish a base point $O_H$ as the origin in ${\cal M}_H$. We choose $O_H=(0,0,0,0,H^{-1})$. The tangent space to ${\cal M}_H$ at $O_H$ is the hyperplane ${\cal M}_o = \{ x\in\mathbb{R}^5:\;x^4=H^{-1} \}$ identified as the four-dimensional Minkowski spacetime to which the dS spacetime can be contracted in the limit $H\rightarrow 0$ (the null-curvature limit).

The corresponding Lie algebra can be realized as the linear span of ten Killing vectors
\begin{eqnarray}\label{tenKv}
K_{\alpha\beta} = x_\alpha \partial_\beta - x_\beta \partial_\alpha .
\end{eqnarray}
It is worth underlining that since there is no global time-like Killing vector in dS spacetime (all isometry generators correspond to rotations or Lorentz boosts), the adjective time-like or space-like is used by referring to the Lorentzian four-dimensional metric induced by that of the ambient spacetime $\mathbb{R}^5$.

The universal covering group of $SO_0(1,4)$ is the (pseudo-)symplectic group $Sp(2,2)$. It is a subgroup of the group of $2\times2$ quaternionic matrices which reads
\begin{eqnarray}
Sp(2,2)=\{g = \left( \begin{array}{ccc}
a & b\\
c & d \end{array} \right):\; \mbox{det} g=1,\; g^\dagger \gamma^0 g=\gamma^0 \},\nonumber
\end{eqnarray}
where $a,b,c,d \in \mathbb{H}$, $g^\dagger = g^{\star t}$ ($g^\star$ being the quaternionic conjugate of $g$ and $g^t$ is the transpose of $g$) and
\begin{eqnarray}
\gamma^0 = \left( \begin{array}{ccc}
\mathbb{I}_{2\times 2} & 0\\
0 & -\mathbb{I}_{2\times 2} \end{array} \right). \nonumber
\end{eqnarray}
Here, ``$\mbox{det} g$" must be understood as a determinant of a fourth-order matrix with complex coefficients resulting from isomorphism $\mathbb{H}\approx \mathbb{R}\times SU(2)$.

The matrix $\gamma^0$ along with the following matrices
\begin{eqnarray}
\gamma^4 = \left( \begin{array}{ccc}
0 & \mathbb{I}_{2\times 2} \\
-\mathbb{I}_{2\times 2} & 0 \end{array}\right), \;
\gamma^k = \left( \begin{array}{ccc}
0 & e_k \\
e_k & 0 \end{array} \right), \; k=1,2,3,\nonumber
\end{eqnarray}
where $e_k =i (-1)^{k+1}\sigma_k$ ($\sigma_k$ being the usual Pauli matrices), satisfy the anticommutation relations $\gamma^\alpha \gamma^\beta + \gamma^\beta \gamma^\alpha = 2\eta^{\alpha\beta}\mathbb{I}_{4\times 4}$ and $\gamma^{\alpha ^\dagger}=\gamma^0\gamma^\alpha\gamma^0$. These matrices are the generating elements of the Clifford algebra based on the metric $\eta_{\alpha\beta}$. There is an isomorphism $SO_0(1,4)\approx Sp(2,2)/\mathbb{Z}_2$ through \cite{Grensing}
\begin{eqnarray}
SO_0(1,4)\ni \Lambda_g:x\longrightarrow \Lambda_g x= x^\prime ,
\end{eqnarray}
where $(\Lambda_g)_\beta^\alpha = \frac{1}{4}\mbox{tr}(\gamma^\alpha g\gamma_\beta g^{-1})$. In Appendix (\ref{factorization}), resorting to a particular (non-global) factorization of the group, we discuss another way of understanding this group action on dS spacetime.

\subsection{$1+3$-dS unitary irreducible representations}
By construction, as pointed out above, dS space is homogeneous and has the large isometry group $SO_0(1,4)$. In this regard, it is very tempting to extend the (Wignerian) group-theoretic ideas, underlying relativistic quantum mechanics and QFT in Minkowski spacetime, to dS spacetime in order to construct dS elementary systems. To achieve this goal, the ambient space notations will be considered here. This way of describing dS spacetime provides a remarkable coordinate-independent approach, such that there is a close resemblance with the corresponding description on Minkowski space. Moreover, within this context the link with group theory would be easily readable.

Let us consider $\psi^{(s)}_{\alpha ...}(x)$ (omitting the spinor index $i$) as a free spinor-tensor field with tensorial rank $n$ and with four spinor components in the ambient space notations ($s=n+\frac{1}{2}$). Note that: (\emph{i}) From now on tensors indices will be dropped whenever possible; (\emph{ii}) The value of $s$ should generally be dissociated from the value of the spin which the latter carries the group-theoretical content of the theory. This point will be amply illustrated in what follows.

In the ambient space framework, the spinor-tensor field $\psi^{(s)}(x)$ is considered as a homogeneous function of the $\mathbb{R}^5$-variables $x^\alpha$, with an arbitrarily chosen homogeneity degree $\varrho$,
\begin{eqnarray}\label{homoge}
x\cdot\partial \psi^{(s)}(x) = \varrho \psi^{(s)}(x),\;\;\;\; x\cdot\partial \equiv x^\alpha \frac{\partial}{\partial x^\alpha}.
\end{eqnarray}
For simplicity reasons, we here set $\varrho=0$. Of course, every homogeneous spinor-tensor field of $\mathbb{R}^5$-variables does not represent a physical dS entity. Indeed, the field also needs to verify the requirement of transversality to ensure that it lies in the dS tangent spacetime,
\begin{eqnarray}\label{transversality}
x\cdot \psi^{(s)}(x) = 0.
\end{eqnarray}
Regarding the significance of this requirement for dS fields, the symmetric, transverse projector $\theta_{\alpha\beta} = \eta_{\alpha\beta} + H^2x_\alpha x_\beta$, verifying $\theta_{\alpha\beta}x^\alpha = \theta_{\alpha\beta}x^\beta = 0$, is defined. Transverse entities are constructed using this transverse projector. For example, the transverse derivative would be $\partial^\top_{\alpha} = \theta_{\alpha\beta} \partial^\beta = \partial_\alpha + H^2 x_\alpha x\cdot\partial$, and we note the properties
\begin{eqnarray}
{\partial}^\top_\alpha x_\beta = \theta_{\alpha\beta},\;\;\; {\partial}^\top_\alpha x^2=0.\nonumber
\end{eqnarray}
Thus, the differential operator ${\partial}^\top$ is intrinsically defined on the hyperboloid $x^2=-H^{-2}$. Generally, for a field with tensorial rank $n$, the transverse projection $\mbox{T}$ with the following definition
\begin{eqnarray}
(\mbox{T}\psi)^{(s)}_{\alpha_1...\alpha_n} = \Big( \prod_{i=1}^n \theta_{\alpha_i}^{\beta_i} \Big) \psi^{(s)}_{\beta_1...\beta_n},
\end{eqnarray}
guarantees the transversality in each index.

We now turn to the description of the UIRs of the dS (universal covering) group $Sp(2,2)$. To do this, we need to introduce self-adjoint operators, one for each of the ten Killing vectors (\ref{tenKv}), in Hilbert space of symmetric spinor-tensors $\psi^{(s)}_{\alpha ...}$ on ${\cal M}_H$, square integrable regarding some invariant inner (Klein-Gordon type) product. These operators can be represented as \cite{Moylan}
\begin{eqnarray}\label{selfad}
K_{\alpha\beta}\longrightarrow L_{\alpha\beta}^{(s)} = L_{\alpha\beta}^{(n)} + S^{(\frac{1}{2})}_{\alpha\beta},
\end{eqnarray}
with
\begin{eqnarray}
S^{(\frac{1}{2})}_{\alpha\beta} = -\frac{i}{4}[\gamma_\alpha,\gamma_\beta]\;\;\;\mbox{and}\;\;\; L_{\alpha\beta}^{(n)} = M_{\alpha\beta}+S_{\alpha\beta}^{(n)},\nonumber
\end{eqnarray}
where $S^{(\frac{1}{2})}_{\alpha\beta}$ acts upon the spinor indices, while $M_{\alpha\beta} = -i(x_\alpha \partial_\beta - x_\beta \partial_\alpha ) = -i(x_\alpha \partial^\top_\beta - x_\beta \partial^\top_\alpha )$ is the orbital part and $S_{\alpha\beta}^{(n)}$ designates the integer spin part. The latter acts on the tensorial indices in the following way
\begin{eqnarray}
S_{\alpha\beta}^{(n)}\psi^{(s)}_{\alpha_1 ... \alpha_n} = -i\sum_{i=1}^{n}(\eta_{\alpha\alpha_i}\psi^{(s)}_{\alpha_1 ...(\alpha_i\rightarrow \beta)... \alpha_n} - (\alpha \rightleftharpoons \beta)).\nonumber
\end{eqnarray}
These generator representatives $L_{\alpha\beta}^{(s)}$ obey the following commutation relations
\begin{eqnarray}
[L_{\alpha\beta}^{(s)}, L_{\gamma\delta}^{(s)}]=-i(\eta_{\alpha\gamma}L_{\beta\delta}^{(s)} +\eta_{\beta\delta}L_{\alpha\gamma}^{(s)} - \eta_{\alpha\delta}L_{\beta\gamma}^{(s)} - \eta_{\beta\gamma}L_{\alpha\delta}^{(s)}).\nonumber
\end{eqnarray}

The second-order dS Casimir operator representative is defined by\footnote{Note that the dS group has two independent Casimir operators: the quadratic one given in (\ref{soCo}) and the quartic one defined by
$$ Q^\prime_s = - W^{(s)}_{\alpha} W^{(s)\alpha}, \;\;\; W^{(s)}_{\alpha}= - \frac{1}{8} \epsilon_{\alpha\beta\gamma\delta\eta} L^{(s)\beta\gamma} L^{(s)\delta\eta}, $$
where $W^{(s)}_\alpha$ is the dS counterpart of the Pauli-Lubanski operator and $ \epsilon_{\alpha\beta\gamma\delta\eta} $ refers to the usual antisymmetrical tensor. In this paper, however, we just focus on the action of the second-order (or quadratic) one.}
\begin{eqnarray}\label{soCo}
Q_s = - \frac{1}{2}L_{\alpha\beta}^{(s)}L^{(s)\;\alpha\beta}.
\end{eqnarray}
It commutes with all generator representatives $L_{\alpha\beta}^{(s)}$. In this sense, $Q_s$ acts as a constant on all states in a given dS UIR, so that
\begin{eqnarray}\label{eigenfieq}
Q_s \psi^{(s)} = \langle Q_s \rangle \psi^{(s)},
\end{eqnarray}
where $\langle Q_s \rangle$ stand for the eigenvalues of $Q_s$. As a consequence, the eigenvalues of $Q_s$ can be used to classify the dS UIRs. Following Dixmier \cite{Dix}, the dS UIRs can be labeled by using a pair of parameters $\Delta=(p,q)$, with $2p\in\mathbb{N}$ and $q\in \mathbb{C}$, in terms of which the eigenvalues of $Q_s$ are
\begin{eqnarray}
\langle Q_s\rangle = [-p(p+1)-(q+1)(q-2)]\mathbb{I}.
\end{eqnarray}
In this sense, regarding the spectral values assumed by the Casimir operator (or the possible values of $p$ and $q$), one can distinguish three series or UIRs of $Sp(2,2)$ as \cite{Dix, Taka} follows:
\begin{itemize}
\item{Principal series representations $U_{s,\nu}$ labelled by $\Delta = (s,\frac{1}{2}+i\nu)$. Here, the parameter $p=s$ has a spin meaning. We must distinguish between:\\
(\emph{i}) the integer spin principal series, with $\nu\in \mathbb{R}$ and $\nu\geq 0$, while $s=0,1,2,...$,\\
(\emph{ii}) the half-integer spin principal series, with $\nu\in \mathbb{R}$ and $\nu > 0$, while $s=\frac{1}{2},\frac{3}{2},\frac{5}{2},...$ .}
\item{Complementary series representations $V_{s,\nu}$ labelled by $\Delta = (s,\frac{1}{2}+\nu)$. Here, the parameter $p=s$ has a spin meaning. We have to distinguish between:\\
(\emph{i}) the scalar case $V_{0,\nu}$, with $\nu\in \mathbb{R}$ and $0<|\nu|<\frac{3}{2}$,\\
(\emph{ii}) the spinorial case $V_{s,\nu}$, with $\nu\in \mathbb{R}$ and $0<|\nu|<\frac{1}{2}$, while $s=1,2,3,...$ .}
\item{Discrete series representations $\Pi_{p,s}^\pm$ labelled by $\Delta = (p,s)$. Unlike the above cases, the parameter $q=s$ has a spin meaning. We have to distinguish between:\\
(\emph{i}) the scalar case $\Pi_{p,0}$, with $p=1,2,...$,\\
(\emph{ii}) the spinorial case $\Pi_{p,s}^\pm$, with $p=\frac{1}{2},1,\frac{3}{2},2,...$ and $s=p,p-1,...,1 \; \mbox{or}\; \frac{1}{2}$. [The physical meaning of the superscript $\pm$ will be discussed in the next subsection.]}
\end{itemize}
Note that in all cases, the corresponding Casimir eigenvalues do not change by the substitution $q\rightarrow (1-q)$. In other words, the representations labeled by $\Delta = (p,q)$ and $\Delta = (p,1-q)$ share same Casimir eigenvalues. By definition, such representations are called Weyl equivalent.

\subsection{Physical interpretation of the dS UIRs}
At this point, it is crucial to understand the physical content of the above representations from the point of view of a Minkowskian observer, i.e., at the contraction limit $H\rightarrow 0$, which clarifies what we mean by massive or massless dS UIRs. Let us recall that the dS metric and the dS isometry group $SO_0(1,4)$ are respectively one parameter deformations of the Minkowski metric and of the proper Poincar$\acute{\mbox{e}}$ group. But, the absence of a global time-like Killing vector field in dS space makes it different from the Minkowski space. In fact, there is no positive conserved energy in dS space. In this sense, besides the above classification of the dS UIRs, we also need to distinguish between those representations of the dS group which contract towards the massive Poincar$\acute{\mbox{e}}$ UIRs and those which have massless content.

On this basis, the massive case only involves the principal series UIR's, i.e., $U_{s,\nu}$. As a matter of fact, if we denote by ${\cal P}^>(s,m)$ and ${\cal P}^<(s,m)$, respectively, the positive and negative energy Wigner UIRs of the Poincar$\acute{\mbox{e}}$ group with mass $m$ and spin $s$, a contraction of the representations $U_{s,\nu}$ is carried out by letting $H\rightarrow 0$ and $\nu\rightarrow \infty$, while the product $\nu H$ remains constant and equals to the Poincar$\acute{\mbox{e}}$ mass $m$, and it yields \cite{Mickelsson, GHR}
\begin{eqnarray}
U_{s,\nu} \longrightarrow {\cal P}^>(s,m) \oplus {\cal P}^<(s,m).\nonumber
\end{eqnarray}
More technically, this contraction is done with respect to the subgroup $SO_0(1,3)$ recognized as the Lorentz subgroup in both Poincar$\acute{\mbox{e}}$ and dS relativities. The associated dS representations, on the other hand, are exactly those representations that are induced by the \emph{minimal parabolic} \cite{Lipsman} subgroup $SO(3)\times SO (1,1) \times\mbox{(a certain nilpotent subgroup)}$, in which $SO(3)$ refers to the space rotation subgroup of the Lorentz subgroup in both cases. This feature completely illuminates the concept of spin in dS spacetime because it comes out of the \emph{same} $SO(3)$.

For massless cases, we select those dS UIRs having a unique extension to the conformal group $SO_0(2,4)$ and that extension would be equivalent to the conformal extension of the massless Poincar$\acute{\mbox{e}}$ UIRs \cite{AFFS, barut}. There are two kinds of these representations:
\begin{itemize}
\item The scalar massless case, which only involves the complementary series UIR $V_{0,\frac{1}{2}}$ labeled by $\Delta = (0,1)$. In the limit $H\rightarrow 0$, it contracts to the massless scalar Poincar$\acute{\mbox{e}}$ UIR,
\begin{eqnarray}
V_{0,\frac{1}{2}}\longrightarrow  {\cal P}^>(0,0) \oplus {\cal P}^<(0,0).\nonumber
\end{eqnarray}
\item The spinorial massless cases, which involve all representations $\Pi_{p,q}^\pm$, with $p=q=s$ ($s>0$), lying at the lower end of the discrete series (for which they are usually called ``dS massless representations"). In the limit $H\rightarrow 0$, they contract to the massless spinorial Poincar$\acute{\mbox{e}}$ UIRs,
\begin{eqnarray}
\Pi_{s,s}^\pm \longrightarrow {\cal P}^>(\pm s,0) \oplus {\cal P}^< (\pm s,0),\nonumber
\end{eqnarray}
here $\pm$ stands for the helicity.
\end{itemize}
Note that all other dS UIRs have either non-physical Poincar$\acute{\mbox{e}}$ contraction limit or have no flat limit at all.

Here, we must underline that although it would seem that labeling as massive or massless dS UIRs, in terms of contraction of representations, could eliminate the interpretative problem of a mass in dS space, it is not the case!. Actually, the notion of mass in ``de Sitterian Physics" still appears ambiguous in terms of contraction of representation, exemplified by the fact that one cannot give a precise meaning to a dS rest energy,\footnote{As we can see from the above argument, the dS representations contract to a direct sum of two UIRs of the Poincar$\acute{\mbox{e}}$ group with positive and negative signs of energy.} except if one follows an approach based on a causality de Sitterian semi-group \cite{Mizony}, or based on a analyticity prerequisite \cite{BrosPRL}: the latter is the way we follow in this paper. Nevertheless, Garidi \cite{What} has proposed a consistent description of de Sitterian mass that depends on the dS UIR's parameters $p$ and $q$. This de Sitterian or Garidi mass is given by
\begin{eqnarray}
m_{H}^2 = \langle Q_s \rangle - \langle Q_s \rangle_{p=q} = [(p-q)(p+q-1)]\hbar^2 H^2/c^4.\nonumber
\end{eqnarray}
Because the lowest value of $\langle Q_s \rangle$ is obtained by setting $p=q$, the above formula implies that every dS UIR, meaningful from the point of view of a Minkowskian observer, gets $m_H^2\geq 0$. The value of the Garidi mass corresponding to the dS UIRs in the principal series is
\begin{eqnarray}
m_H^2= [\nu^2+(s-1/2)^2]\hbar^2 H^2/c^4,\nonumber
\end{eqnarray}
and in the complementary series is
\begin{eqnarray}
m_H^2=[(s-1/2)^2-\nu^2]\hbar^2 H^2/c^4.\nonumber
\end{eqnarray}
Note that $\nu$ is known as the mass parameter when we fix $s$ to one of its possible values.

\section{dS Rarita-Schwinger equation}
We are now in a position to study the dS massive spin-$\frac{3}{2}$ field $\psi_\alpha^{(\frac{3}{2})}$, with tensorial rank $n=1$. According to the above discussion, it is immediately seen that this field corresponds to a specific UIR of the principal series, i.e., $U_{\frac{3}{2}, \nu}$, labeled here by $\Delta = (\frac{3}{2},\frac{1}{2} + i\nu)$. In this sense, the most familiar way to characterize the corresponding dS group representation and the carrier states space would be to consider the solutions to the ``wave equation" (\ref{eigenfieq}), with $s=\frac{3}{2}$ and $\langle Q_{\frac{3}{2}}\rangle = \nu^2 - \frac{3}{2}$, that is,
\begin{eqnarray}\label{Field Eq.}
\Big(Q_{\frac{3}{2}} - \nu^2 + \frac{3}{2}\Big)\psi^{(\frac{3}{2})} = 0.
\end{eqnarray}

To determine the solutions to the above equation, we write the action of $Q_\frac{3}{2}$ on $\psi^{(\frac{3}{2})}$ in the following explicit form
\begin{eqnarray}\label{apQtopsi}
Q_\frac{3}{2}\psi^{(\frac{3}{2})} &=& \Big(Q_{\frac{1}{2}} - 3\Big)\psi^{(\frac{3}{2})} -2 \partial x\cdot \psi^{(\frac{3}{2})} + 2 x\partial\cdot\psi^{(\frac{3}{2})} \nonumber\\
&& + \gamma (\gamma\cdot\psi^{(\frac{3}{2})}),
\end{eqnarray}
in which
\begin{eqnarray}\label{Q1-2op}
Q_{\frac{1}{2}} = Q_0 + \frac{i}{2}\gamma_\alpha\gamma_\beta M^{\alpha\beta} - \frac{5}{2},
\end{eqnarray}
and $Q_0\equiv -\frac{1}{2} M_{\alpha\beta}M^{\alpha\beta}$ is the scalar Casimir operator. Considering Eq. (\ref{apQtopsi}), it is obvious that the space of solutions to (\ref{Field Eq.}) contains some invariant subspaces that must be eliminated if one wishes to be left with a space that carries $U_{\frac{3}{2}, \nu}$. Along with the aforementioned homogeneity and transversality conditions, this is done by imposing the supplementary conditions of divergencelessness and tracelessness. We list these supplementary conditions, respectively, below
\begin{eqnarray}\label{suu}
x\cdot\partial \psi^{(\frac{3}{2})}&=&0,\nonumber\\
x\cdot\psi^{(\frac{3}{2})}&=&0,\nonumber\\
\partial\cdot\psi^{(\frac{3}{2})}&=&0,\nonumber\\
\gamma\cdot\psi^{(\frac{3}{2})}&=&0.
\end{eqnarray}
Having Eq. (\ref{apQtopsi}), the above conditions permit one to rewrite Eq. (\ref{Field Eq.}) as
\begin{eqnarray}\label{divergencelessness}
\Big(Q_{\frac{1}{2}} - \nu^2 - \frac{3}{2}\Big)\psi^{({\frac{3}{2}})} = 0.
\end{eqnarray}

On the other hand, using the following identity
\begin{eqnarray}\label{identity}
\Big( \frac{1}{2}\gamma_\alpha\gamma_\beta M^{\alpha\beta}\Big)^2 =  Q_0 - \frac{3i}{2}\gamma_\alpha\gamma_\beta M^{\alpha\beta},
\end{eqnarray}
one can rewrite (\ref{Q1-2op}) as
\begin{eqnarray}\label{Q_1/2 new}
Q_{\frac{1}{2}} = \Big( \frac{1}{2}\gamma_\alpha\gamma_\beta M^{\alpha\beta} + 2i \Big)^2 + \frac{3}{2}.
\end{eqnarray}
Consequently, we can express Eq. (\ref{divergencelessness}) as
\begin{eqnarray}\label{Field Eq. new}
\Big( \kappa -2-i\nu \Big)\Big( \kappa -2+i\nu \Big)\psi^{({\frac{3}{2}})} = 0,
\end{eqnarray}
where $\kappa\equiv \frac{i}{2}\gamma_\alpha\gamma_\beta M^{\alpha\beta} = \slashed{x} \slashed{\partial}^\top$ ($\slashed{x}$ and $\slashed{\partial}^\top$ denote $\gamma\cdot x$ and $\gamma\cdot\partial^\top$, respectively) is the Dirac operator. Now, recalling that theories corresponding to $U_{s,\nu}$ and $U_{s,-\nu}$ are indeed equivalent (see the previous section), one obtains a considerable simplification in realizing the space of solutions to Eq. (\ref{Field Eq.}) as the space of solutions to the following first-order equation
\begin{eqnarray}\label{Field Eq. newwwwww}
\Big( \kappa -2 + i\nu \Big)\psi^{({\frac{3}{2}})} = 0.
\end{eqnarray}
It is worth pointing out that the above equation corresponds to the usual Rarita-Schwinger equation in curved spacetime written in terms of covariant derivative (see \cite{FaFr} for more details). Then, we shall call it the dS Rarita-Schwinger first-order equation.

The adjoint spinor-vector $\overline{{\psi}}^{({\frac{3}{2}})}$ is related to its Hermitian conjugate ${\psi^{({\frac{3}{2}})}}^\dagger$ by
\begin{eqnarray}\label{adjoint}
\overline{{\psi}}^{({\frac{3}{2}})}(x) = {\psi^{({\frac{3}{2}})}}^\dagger(x) \gamma^0\gamma^4.
\end{eqnarray}
It verifies the following adjoint equation
\begin{eqnarray} \label{Eq. adjoint}
\overline{{\psi}}^{({\frac{3}{2}})}\Big[-\gamma^4\Big( \overleftarrow{\kappa} -2-i\nu \Big)\gamma^4\Big] = 0,
\end{eqnarray}
in which, considering the conventional notation, $\overleftarrow{\kappa}$ means that the derivatives act to the left.

The transformation rules for ${\psi^{({\frac{3}{2}})}}(x)$ and its adjoint $\overline{{\psi}}^{({\frac{3}{2}})}(x)$, based upon which Eqs. (\ref{Field Eq. newwwwww}) and (\ref{Eq. adjoint}) are covariant, would be
\begin{eqnarray} \label{covariance}
{\psi^{({\frac{3}{2}})}}(x) &\rightarrow& {{\psi^\prime}^{({\frac{3}{2}})}}(x) = g^{-1} \psi (\Lambda_gx),\nonumber\\
\overline{{\psi}}^{({\frac{3}{2}})}(x) &\rightarrow& {\overline{{\psi}}^\prime}^{({\frac{3}{2}})}(x) = \overline{{\psi}}^{({\frac{3}{2}})}(\Lambda_gx)i(g),
\end{eqnarray}
where $g$ belongs to the (pseudo-)symplectic group $Sp(2,2)$ and $i(g) = \gamma^0\gamma^4 g^\dagger \gamma^0\gamma^4$ is a group involution in $Sp(2,2)$ \cite{Taka}.

\section{dS spinor-vector waves}

\subsection{Field equation solutions}
In this section, we produce a recurrence formula allowing one to give the general solutions to the dS Rarita-Schwinger first-order equation (\ref{Field Eq. newwwwww}). This recurrence formula involves operators which make the spinor-vector field $\psi^{(\frac{3}{2})}$ of tensorial rank $n=1$ from a spinor field $\psi^{(\frac{1}{2})}$ of tensorial rank $n=0$. Such operators are truly expected to obey commutation/intertwining rules with $L_{\alpha\beta}^{(\frac{3}{2})}$ and $Q_\frac{3}{2}$ (regarding the above simplification, say $\kappa$).

In such a recurrence formula, the contraction of the transverse projector $\theta$ with a constant polarization five-vector ${A}$ (i.e., $\theta\cdot {A} = {A}^\top$), which permits one to define an operator that makes a transverse spinor-vector $\psi^{(\frac{3}{2})}$ from an arbitrary spinor field $\chi$, can be considered as an ingredient part. In this sense, let us first determine the commutation relation between $\kappa$ and ${A}^\top$. It takes the form
\begin{eqnarray}\label{first commutation}
[\kappa,\; {A}^\top] \chi = - H^2 {\gamma}^\top\slashed{x}(x\cdot {A})\chi,
\end{eqnarray}
in which $\gamma^\top = \theta\cdot\gamma$. Considering the above relation, now we need to obtain the commutation relation between $\kappa$ and the new element $\gamma^\top$. Let $\chi$ be an arbitrary spinor field, then for this new element one finds
\begin{eqnarray}\label{second commutation}
[\kappa,\; \gamma^\top] \chi =  2H^2 D_1 \slashed{x}\chi - \gamma^\top \chi,
\end{eqnarray}
where $D_1 = H^{-2}\partial^\top$ is the generalized gradient associated with the tensorial part on the dS hyperboloid \cite{Gazeau88}, for which, we have $Q_1D_1=D_1Q_0$. The definition of the generalized gradient can be extended to the generalized gradient on the dS hyperboloid for spinor-vectors through $D_{\frac{3}{2}}=D_1 + \gamma^\top\slashed{x}$, for which, the commutation rule $Q_{\frac{3}{2}}D_{\frac{3}{2}} = D_{\frac{3}{2}}Q_{\frac{1}{2}}$ holds. With this extended definition of the dS gradient, the relation (\ref{second commutation}) can be rewritten as\footnote{Note that $\slashed{x}\slashed{x}=-H^{-2}$.}
\begin{eqnarray}\label{second commutation2}
[\kappa,\; \gamma^\top] \chi = 2H^2 D_{\frac{3}{2}} \slashed{x}\chi + \gamma^\top \chi.
\end{eqnarray}
This relation (\ref{second commutation2}) implies that we need to evaluate the commutator between the new element $D_{\frac{3}{2}}$ and $\kappa$, as well. Again, considering $\chi$ as an arbitrary spinor field, we have
\begin{eqnarray}\label{third commutation}
[\kappa,\; D_{\frac{3}{2}}]\chi &=& - H^{-2} {\gamma^\top}{\slashed\partial}^\top \chi - D_{\frac{3}{2}} \chi + 4{\gamma^\top}\slashed{x}\chi \nonumber\\
&& -2 {\gamma^\top}\slashed{x} \kappa\chi.
\end{eqnarray}

The above commutation relations explicitly reveal that under the action of $\kappa$ (say the action of $Q_{\frac{3}{2}}$) all the appeared terms are of the form ${A}^\top\chi$, $\gamma^\top\varsigma$ and $D_{\frac{3}{2}}\rho$, where $\chi$, $\varsigma$ and $\rho$ are spinor fields of tensorial rank $n=0$. On this basis, the spinor-vector field $\psi^{(\frac{3}{2})}$, as the general solutions to (\ref{Field Eq. newwwwww}) and (\ref{suu}), can be put in the following dS-invariant recurrence form
\begin{eqnarray} \label{solu 1}
\psi^{(\frac{3}{2})} = {{A}^\top}\psi^{(\frac{1}{2})} + D_{\frac{3}{2}}\psi^{(\frac{1}{2})}_2 + {\gamma^\top}\psi^{(\frac{1}{2})}_3,
\end{eqnarray}
in which the spinors $\psi^{(\frac{1}{2})}$, $\psi^{(\frac{1}{2})}_2$ and $\psi^{(\frac{1}{2})}_3$ need to explicitly be determined.

Substituting the above general solutions (\ref{solu 1}) into the field equation (\ref{Field Eq. newwwwww}) and after a direct calculation, one finds from the linear independence of the terms in (\ref{solu 1}) that
\begin{eqnarray} \label{Field Eq. 01}
\Big( \kappa + N \Big)\psi^{(\frac{1}{2})} = 0,
\end{eqnarray}
\begin{eqnarray} \label{Field Eq. 02}
\Big( \kappa + N -1 \Big)\psi^{(\frac{1}{2})}_2 + 2H^2\slashed{x}\psi^{(\frac{1}{2})}_3  = 0,
\end{eqnarray}
\begin{eqnarray} \label{Field Eq. 03}
\Big( \kappa + N -1 \Big)\psi^{(\frac{1}{2})}_3 + \slashed{x}\Big( 3 + N \Big)\psi^{(\frac{1}{2})}_2 \hspace{2cm}\nonumber\\
- H^2\slashed{x}(x\cdot {A})\psi^{(\frac{1}{2})} = 0,
\end{eqnarray}
where, for the sake of simplicity of the notation, we define $N \equiv -2+i\nu$. It is interesting to note that Eq. (\ref{Field Eq. 01}) corresponds to the usual Dirac equation in curved spacetime written in terms of covariant derivative \cite{Diracfields}. In this sense, and following Bartesaghi et al \cite{Diracfields}, we call it the dS Dirac equation. In what follows, we shall show that the spinors $\psi^{(\frac{1}{2})}_2$ and $\psi^{(\frac{1}{2})}_3$ are completely determined in terms of $\psi^{(\frac{1}{2})}$.

Let us begin with $\psi^{(\frac{1}{2})}_2$. Imposing the tracelessness condition ($\gamma\cdot\psi^{(\frac{3}{2})}=0$) on the recurrence formula (\ref{solu 1}), one obtains
\begin{eqnarray} \label{tracelessness new}
\psi_{3}^{(\frac{1}{2})} = \frac{-1}{4} \Big( {\slashed{{A}}}^\top\psi^{(\frac{1}{2})} + H^{-2}{\slashed\partial}^\top \psi_{2}^{(\frac{1}{2})} + 4\slashed{x}\psi_{2}^{(\frac{1}{2})}\Big),
\end{eqnarray}
where ${\slashed{{A}}}^\top = \gamma\cdot{{A}^\top}$. Combining (\ref{Field Eq. 02}) with (\ref{tracelessness new}), we have
\begin{eqnarray} \label{Field Eq. 02new}
\Big( \kappa +2N +2 \Big)\psi_{2}^{(\frac{1}{2})} + H^2{\slashed{{A}}}^\top\slashed{x}\psi^{(\frac{1}{2})}  = 0.
\end{eqnarray}
The general form of $\psi_{2}^{(\frac{1}{2})}$ then can be written in terms of $\psi^{(\frac{1}{2})}$ as follows
\begin{eqnarray} \label{2-->1}
\psi_{2}^{(\frac{1}{2})} &=& aH^2(x\cdot {A}) \psi^{(\frac{1}{2})} + b({A}\cdot{\partial^\top})\psi^{(\frac{1}{2})} + cH^2{\slashed{{A}}}^\top\slashed{x}\psi^{(\frac{1}{2})}\nonumber\\
&\equiv& [a,b,c].
\end{eqnarray}
where $[a,b,c]$ is an element of the three-dimensional space $E$ generated by three basic functions $H^2(x\cdot {A})\psi^{(\frac{1}{2})}$, $({A}\cdot{\partial^\top})\psi^{(\frac{1}{2})}$ and $H^2{\slashed{{A}}}^\top\slashed{x}\psi^{(\frac{1}{2})}$. In fact, by considering the last term in (\ref{Field Eq. 02new}) as a guideline, we choose these functions such that $[a,b,c]$ be invariant under the action of $\kappa$,
\begin{eqnarray}\label{inka1}
\kappa H^2(x\cdot {A}) \psi^{(\frac{1}{2})} &=& [-N,0,-1],\hspace{2.2cm}
\end{eqnarray}
\begin{eqnarray}\label{inka2}
\kappa ({A}\cdot{\partial^\top})\psi^{(\frac{1}{2})} = [-N,-N+1,-N],\;\;\;\;
\end{eqnarray}
\begin{eqnarray}\label{inka3}
\kappa H^2{\slashed{{A}}}^\top\slashed{x} \psi^{(\frac{1}{2})} = [-4,-2,N+3].\hspace{0.3cm}\;\;\;\;\;\;
\end{eqnarray}
After inserting (\ref{2-->1}) into (\ref{Field Eq. 02new}) and with the help of (\ref{inka1}), (\ref{inka2}) and (\ref{inka3}), one can easily find the following system
\begin{eqnarray}
\left(\begin{array}{ccc} 2+N &  -N & -4 \\ 0 & 3+N & -2 \\ -1 & -N & 5+3N \\
\end{array}\right) \left(\begin{array}{ccc} a \\ b \\ c \\ \end{array}\right)= \left(\begin{array}{ccc} 0 \\ 0 \\ -1 \\ \end{array}\right).
\end{eqnarray}
Note that the determinant of the coefficient matrix is non-zero. Then, the expression for $\psi_{2}^{(\frac{1}{2})}$ in (\ref{Field Eq. 02new}) is obtained by solving the above system, which results in
\begin{eqnarray} \label{2-->1 finished}
\psi_{2}^{(\frac{1}{2})} = \frac{-1}{3(N+1)}\Big[\frac{6}{N+3},\frac{2}{N+3},1\Big].
\end{eqnarray}

We are now in the position to determine $\psi_{3}^{(\frac{1}{2})}$ in terms of $\psi^{(\frac{1}{2})}$. Considering (\ref{2-->1 finished}), the inhomogeneous Eq. (\ref{Field Eq. 03}) can be written as
\begin{eqnarray} \label{3-->1}
\Big( \kappa + N -1 \Big)\psi_{3}^{(\frac{1}{2})} =\hspace{4cm}\nonumber\\
\frac{1}{3(N+1)}\{3N+9,2,N+3\},
\end{eqnarray}
where
\begin{eqnarray} \label{Field Eq. 03new}
\{e,f,g\} &=& e H^2\slashed{x}(x\cdot {A})\psi^{(\frac{1}{2})} + f\slashed{x}({A}\cdot{\partial^\top})\psi^{(\frac{1}{2})} \nonumber\\
&&+ g H^2\slashed{x}{\slashed{{A}}}^\top\slashed{x}\psi^{(\frac{1}{2})},
\end{eqnarray}
is an element of another three-dimensional space $E^\prime$ generated by three basic functions $H^2\slashed{x}(x\cdot {A})\psi^{(\frac{1}{2})}$, $\slashed{x}({A}\cdot{\partial^\top})\psi^{(\frac{1}{2})}$ and $H^2\slashed{x}{\slashed{{A}}}^\top\slashed{x}\psi^{(\frac{1}{2})}$. This space is invariant under the action of $\kappa$,
\begin{eqnarray}
\kappa H^2\slashed{x}(x\cdot {A}) \psi^{(\frac{1}{2})} = \{4+N,0,1\},\;\;\;\;\;\;
\end{eqnarray}
\begin{eqnarray}
\kappa \slashed{x}({A}\cdot{\partial^\top})\psi^{(\frac{1}{2})} = \{N,3+N,N\},\;
\end{eqnarray}
\begin{eqnarray}
\kappa H^2\slashed{x}{\slashed{{A}}}^\top\slashed{x} \psi^{(\frac{1}{2})} = \{4,2,-N+1\}.
\end{eqnarray}
Now, the solution to Eq. (\ref{3-->1}) inside $E^\prime$ can be simply obtained by solving the following system
\begin{eqnarray}
\left(\begin{array}{ccc} 3+2N &  N & 4 \\ 0 & 2+ 2N & 2 \\ 1 & N & 0 \\
\end{array}\right) \left(\begin{array}{ccc} e \\ f \\ g \\ \end{array}\right)=\hspace{3cm}\nonumber\\
\frac{1}{3(N+1)}\left(\begin{array}{ccc} 3N+9 \\ 2 \\ N+3 \\ \end{array}\right).\;\;
\end{eqnarray}
Again, the determinant of the coefficient matrix of the above system is non-zero. Therefore, we have
\begin{eqnarray} \label{3-->1 finished}
\psi_{3}^{(\frac{1}{2})} = \frac{1}{3(N+1)}\{3,1,-N\}.
\end{eqnarray}

Taking all of the above into consideration, the general solutions (\ref{solu 1}) can be written as the resulting action of a dS-invariant differential operator on the spinor field $\psi^{(\frac{1}{2})}$,
\begin{eqnarray} \label{general solu}
&\psi^{(\frac{3}{2})}_\alpha = {{A}_\alpha^\top} \psi^{(\frac{1}{2})}\hspace{5.9cm}& \nonumber\\
&+ \frac{1}{3(N+1)} D_{\frac{3}{2}\alpha}\Big(\frac{-6H^2}{N+3}(x\cdot{{A}}) - \frac{2}{N+3}({A}\cdot{\partial^\top}) &\nonumber\\
&- H^2{\slashed{{A}}}^\top\slashed{x}\Big)\psi^{(\frac{1}{2})}& \nonumber\\
&+ \frac{1}{3(N+1)} {\gamma_\alpha^\top}\Big(3H^2\slashed{x}(x\cdot{{A}}) + \slashed{x}({A}\cdot{\partial^\top})\hspace{0.7cm}\;&\nonumber\\
&- N H^2\slashed{x}{\slashed{{A}}}^\top\slashed{x}\Big) \psi^{(\frac{1}{2})}.&
\end{eqnarray}
Note that these solutions are formally similar to the ones introduced by Lesimple for the massive spin-$\frac{3}{2}$ field in anti-dS spacetime \cite{Lesimple}.

The preceding discussion was an attempt to write the general solutions (\ref{solu 1}) in terms of the spinor $\psi^{(\frac{1}{2})}$ verifying the dS Dirac equation (\ref{Field Eq. 01}). In the following, we shall produce a family of dS Dirac plane waves in order to set up a momentum space representation for the dS Rarita-Schwinger field (\ref{general solu}).

From the factorization of the scalar operator $(\kappa + N)(-\kappa +3 + N) = [Q_0 + N(3+N)]$,\footnote{Here, the identity (\ref{identity}) has been used.} the spinor field $\psi^{(\frac{1}{2})}$ can be obtained from
\begin{eqnarray} \label{1-->0}
\psi^{(\frac{1}{2})} (x)= (-\kappa +3 +N)[\phi(x)v],
\end{eqnarray}
in which $v$ is an arbitrary four-component spinor and $\phi(x)$ is a scalar field obeying the dS Klein-Gordon equation,
\begin{eqnarray}\label{KG equation}
(Q_0 + \mu^2)\phi(x) = 0,
\end{eqnarray}
with $\mu^2 = N(3+N)$ and $-H^2Q_0=\Box_H$ ($\Box_H$ being the Laplace-Beltrami operator on ${\cal M}_H$). In this sense, we can easily fulfill the task of finding the explicit form of $\psi^{(\frac{1}{2})}$ by using the known results for the dS Klein-Gordon equation (\ref{KG equation}).

Let $\xi$ be a null vector in the ambient space, $\xi\in {\cal C} =\{\xi \in \mathbb{R}^5:\;\xi^2=0\}$, then for $x\in {\cal M}_H$ such that $x\cdot\xi\neq 0$, there exists a continuous family of coordinate-independent solutions (``dS plane waves") to Eq. (\ref{KG equation}) (see the details in Ref. \cite{GelV})
\begin{eqnarray}\label{coindSpw}
\phi (x) = (Hx\cdot\xi)^\sigma,\;\;\; \sigma\in\mathbb{C}.
\end{eqnarray}
These solutions are homogeneous with degree $\sigma$ on the null cone ${\cal C}$ and thus are completely determined by specifying their values on a well chosen three dimensional submanifold (the orbital basis) $\Gamma$ of ${\cal C}$. By substituting (\ref{coindSpw}) into (\ref{KG equation}), one finds two possible values of $\sigma$, i.e.,
\begin{eqnarray}\label{sigma}
N \;\;\; \mbox{and} \;\;\; -N-3.
\end{eqnarray}
Correspondingly, the two possible solutions for $\psi^{(\frac{1}{2})}$, see (\ref{1-->0}), would be
\begin{eqnarray}\label{sigma}
\psi^{(\frac{1}{2})}_1(x) &=& - HN(Hx\cdot\xi)^{N-1} \slashed{x}\slashed{\xi}v_1(\xi) \nonumber\\
&& + (3+2N) (Hx\cdot\xi)^Nv_1(\xi),\\
\psi^{(\frac{1}{2})}_2(x) &=& H(N+3)(Hx\cdot\xi)^{-N-4} \slashed{x}\slashed{\xi}v_2(\xi).\;\;\;\;\;\;
\end{eqnarray}

Regarding the identity $\slashed{\xi}^2 = \xi^2 = 0$, the spinor $u_2(\xi)\equiv \slashed{\xi}v_2(\xi)$ obeys the following linear homogeneous equation
\begin{eqnarray}\label{xi xi =0}
\slashed{\xi} u(\xi) = 0.
\end{eqnarray}
Correspondence with the Minkowskian case (which will be discussed in the next subsection) implies that the spinor $u_1(\xi)\equiv v_1(\xi)$ obeys (\ref{xi xi =0}), as well.

Eq. (\ref{xi xi =0}) has two linearly independent solutions, because the rank of the matrix $\slashed{\xi}$ is two $(\mbox{det}(\slashed{\xi}) = (\xi^2)^2=0)$. Accordingly, the dS Dirac equation (\ref{Field Eq. 01}) has four independent solutions \cite{Diracfields}
\begin{eqnarray}\label{Dirac solu 111}
\psi^{(\frac{1}{2})}_{1,r}(x) = (Hx\cdot\xi)^N u_{1,r}(\xi),\hspace{0.8cm}
\end{eqnarray}
\begin{eqnarray}\label{Dirac solu 222}
\psi^{(\frac{1}{2})}_{2,r}(x) = (Hx\cdot\xi)^{-N-4} \slashed{x} u_{2,r}(\xi),
\end{eqnarray}
with $r=1,2$. One may also rewrite them in the following more convenient notation
\begin{eqnarray}\label{Dirac solu 1}
\psi^{(\frac{1}{2})}_{1,r}(x) = (Hx\cdot\xi)^{\sigma_-} u_{1,r}(\xi),\;\;
\end{eqnarray}
\begin{eqnarray}\label{Dirac solu 2}
\psi^{(\frac{1}{2})}_{2,r}(x) = (Hx\cdot\xi)^{\sigma_+} \slashed{x} u_{2,r}(\xi),
\end{eqnarray}
where $\sigma_\mp = -2\pm i\nu$ (recall $N\equiv -2+i\nu$). Note that, considering $\sigma_\mp^\ast$ as complex conjugates of $\sigma_\mp$, we have $\sigma_\mp^\ast = \sigma_\pm$.

By substituting (\ref{Dirac solu 1}) and (\ref{Dirac solu 2}) into (\ref{general solu}), the general solutions (\ref{general solu}) to the dS Rarita-Schwinger equation (\ref{Field Eq. newwwwww}) can be written as
\begin{eqnarray} \label{general solu 1}
\Big({\psi^{(\frac{3}{2})}_{1,r}}\Big)_\alpha(x) = f{\cal E}_{\alpha}(x,\xi,{A}) (Hx\cdot\xi)^{\sigma_-} u_{1,r}(\xi),\;\;\;\;\;
\end{eqnarray}
\begin{eqnarray} \label{general solu 2}
\Big({\psi^{(\frac{3}{2})}_{2,r}}\Big)_\alpha(x) = f^\ast{\cal E}_{\alpha}(x,\xi,{A}) (Hx\cdot\xi)^{\sigma_+} \slashed{x} u_{2,r}(\xi),\;\;
\end{eqnarray}
in which $f={\sigma_-}/({\sigma_-}+1)$ and
\begin{eqnarray} \label{E 1}
{\cal E}_{\alpha}(x,\xi,{A}) = A_\alpha^\top - \frac{\xi_\alpha^\top}{x\cdot\xi}(x\cdot A).
\end{eqnarray}
Note that: (\emph{i}) The above formulas have been found with the choices $A\cdot\xi=0=\gamma\cdot A$, for which, the calculation leads to the simplest form of the polarization vectors ${\cal E}_{\alpha}$ compatible with the Minkowski polarization vectors in the flat limit $H\rightarrow 0$ (we will discuss this matter in more detail in the next subsection); (\emph{ii}) Contrary to the flat space case, the polarization vectors ${\cal E}_{\alpha}$ are functions of spacetime; (\emph{iii}) We have $\xi^\top\cdot {\cal E}=A\cdot\xi =0$, since $\xi^\top\cdot\xi^\top=(Hx\cdot\xi)^2$; (\emph{iv}) One can easily check that the spinor-vector waves (\ref{general solu 1}) and (\ref{general solu 2}) fulfill the auxiliary conditions (\ref{suu}).

On the other hand, the adjoint spinor-vector waves $\overline{\psi}^{\;(\frac{3}{2})}$ satisfying the adjoint dS Rarita-Schwinger equation (\ref{Eq. adjoint}) are given by
\begin{eqnarray} \label{adj general solu 1}
\Big({\overline{\psi}^{\;(\frac{3}{2})}_{1,r}}\Big)_\alpha(x) = f^\ast{{\cal E}}_{\alpha}(x,\xi,{A})(Hx\cdot\xi)^{\sigma_+} \overline{u}_{1,r}(\xi),\;\;\;
\end{eqnarray}
\begin{eqnarray} \label{adj general solu 2}
\Big({\overline{\psi}^{\;(\frac{3}{2})}_{2,r}}\Big)_\alpha(x) = f{{\cal E}}_{\alpha}(x,\xi,{A}) (Hx\cdot\xi)^{\sigma_-} \overline{u}_{2,r}(\xi) \overline{\slashed{x}},\;\;\;
\end{eqnarray}
in which the adjoint spinor is defined by $\overline{u}_r = {u}_r^\dagger \gamma^0\gamma^4$ and $\overline{\slashed{x}}= \gamma^0\gamma^4{\slashed{x}}^\dagger\gamma^0\gamma^4$.

\subsection{Flat limit and analytic spinor-vector waves}
The interpretation of the presented dS spinor-vector waves $\psi^{(\frac{3}{2})}_\alpha(x)$ is made possible by examining their zero-curvature limit. To do this, we use the notion of orbital basis $\Gamma$ for the future null cone ${\cal C}^+=\{\xi\in {\cal C}:\; \mbox{sgn}\;\xi^0=+\}$ with respect to a subgroup $H_e$ of the dS relativity group $SO_0(1,4)$ \cite{Bros}. Here, $H_e$ is the stabilizer of a unit vector $e$ ($|e^2|=1$) in $\mathbb{R}^5$. There are two interesting types of orbits in this context. The first one is the spherical type $\Gamma_0$ associated with $e\in {V^+}$ ($V^+$ being the interior of $\overline{V^+}$ introduced in (\ref{V^+})), which is an orbit of $H_e\approx SO(4)$,\footnote{See more mathematical details in Appendix (\ref{realization}).}
\begin{eqnarray}
\Gamma_0 = \{\xi :\;e\cdot\xi = a>0 \}\cap {\cal C}^+.\nonumber
\end{eqnarray}
However, because the irreducible representation associated with our study admits a massive Poincar$\mbox{\'{e}}$ group UIR in the limit $H\rightarrow 0$, it would be most suitable to restrict $\xi$ to the hyperbolical orbital basis $\Gamma_4$ (the second type) corresponding to $e^2 = -1$, i.e., the union of two hyperboloid sheets that are orbits of the subgroup $H_e\approx SO_0(1,3)$,
\begin{eqnarray}
\Gamma_4 = \{\xi\in{\cal C}^+:\;\xi^{(4)}=1\}\cup\{\xi\in{\cal C}^+:\;\xi^{(4)}=-1\},\nonumber
\end{eqnarray}
and it would also be suitable to parametrize $\xi$ by the wave vector $(k^0,\vec{k})$ of a Minkowskian particle of mass $m$, i.e.,
\begin{eqnarray}\label{paraxi}
\xi_\pm = \Big(\frac{k^0}{mc}=\sqrt{\frac{\vec{k}^2}{m^2c^2}+1}\;,\frac{\vec{k}}{mc}\;,\pm 1\Big).
\end{eqnarray}
We also consider global coordinates given by
\begin{eqnarray}\label{4.3}
\left \{ \begin{array}{rl} x^0 &= H^{-1}\sinh (HX^0),\vspace{2mm}\\
\vspace{2mm} \vec{x} &= (H\|\vec{X}\|)^{-1}\vec{X}\cosh (HX^0)\sin (H\|\vec{X}\|),\\
\vspace{2mm} x^4 &= H^{-1}\cosh (HX^0)\cos (H\|\vec{X}\|),\end{array}\right.
\end{eqnarray}
where the dS point $x=x_H(X)$ is parametrized by the Minkowskian variables $X=(X_0=ct,\vec{X})$. The latter are measured in units of the dS radius $H^{-1}$.

As discussed in the previous subsection, the dS spinor-vector waves are made up of three parts, namely, the polarization vectors, the dS plane waves and the spinors. Here, in order to simplify the analysis, we study the behavior of them in the limit $H\rightarrow 0$, separately. Let us start with the polarization vectors.

\textbf{\emph{Part I.}} For the polarization vectors, we have\footnote{Note that $\lim_{H\rightarrow 0} \theta_{\alpha\beta} = \eta_{\mu\nu}$ and $\lim_{H\rightarrow 0} \xi_\alpha^\top = \frac{k_\mu}{m},\;\; \forall\xi\in\Gamma_4$.} \cite{massive/2}
\begin{eqnarray}\label{poveclimit}
\lim_{H\rightarrow 0} [f{\cal E}_\alpha(x,\xi,A)] = A^{(\lambda)}_\mu - \frac{A^{(\lambda)}_4}{\xi_4}\xi_\mu \equiv \epsilon_\mu^{(\lambda)},
\end{eqnarray}
where $\epsilon_\mu^{(\lambda)}$, with $\mu = 0,1,2,3$ and $\lambda = 1,2,3$, are the three Minkowski polarization four-vectors satisfying the usual relations
\begin{eqnarray}\label{mikpolre}
\epsilon^{(\lambda)}\cdot k&=&0,\nonumber\\
\sum_{\lambda=1}^3 \epsilon_\mu^{(\lambda)} (k) \epsilon_\nu^{(\lambda)} (k) &=& -\Big(\eta_{\mu\nu} - \frac{k_\mu k_\nu}{m^2}\Big),\\
\epsilon^{(\lambda)}\cdot\epsilon^{{(\lambda^\prime)}}&=&\eta^{\lambda\lambda^\prime}.\hspace{0.3cm}\nonumber
\end{eqnarray}
The above relations are fulfilled if the $A^{(\lambda)}$'s are such that
\begin{eqnarray}\label{3.17}
A^{(\lambda)}\cdot \xi &=& 0, \nonumber\\
\sum_{\lambda =1}^3 A_\alpha^{(\lambda)} A_\beta^{(\lambda)} &=& -\eta_{\alpha\beta}, \\
A^{(\lambda)}\cdot A^{{(\lambda^\prime)}}&=&\eta^{\lambda\lambda^\prime},\nonumber\\
\sum_{\lambda =1}^3 A_4^{(\lambda)} A_\mu^{(\lambda)} &=& 0,\;\;\;\forall \mu.\nonumber
\end{eqnarray}

With the help of the relations (\ref{3.17}), we have the following properties of the dS polarization vectors in the ambient space formalism
\begin{eqnarray}
\sum_{\lambda =1}^3 {\cal E}_\alpha^{(\lambda)} (x,\xi,A) {\cal E}_\beta^{(\lambda)} (x,\xi,A)&=& -\Big(\theta_{\alpha\beta} - \frac{\xi_\alpha^\top\xi_\beta^\top}{(Hx\cdot\xi)^2}\Big),\nonumber\\
{\cal E}^{(\lambda)} (x,\xi,A)\cdot {\cal E}^{{(\lambda^\prime)}} (x,\xi,A)&=& \eta^{\lambda\lambda^\prime},\nonumber\\
A^{(\lambda)}\cdot {\cal E}^{{(\lambda^\prime)}}(x,\xi,A) &=& \eta^{\lambda\lambda^\prime},\nonumber\\
{\cal E}^{(\lambda)} (x,\xi,A)\cdot {\cal E}^{{(\lambda^\prime)}} (x^\prime,\xi,A)&=&\eta^{\lambda\lambda^\prime},\nonumber
\end{eqnarray}
which are interestingly similar to the Minkowskian case (\ref{mikpolre}). In this sense, the dS generalized polarization vectors would read
\begin{eqnarray} \label{E 1'}
{\cal E}^{{(\lambda)}}_{\alpha}(x,\xi) \equiv {\cal E}_{\alpha}(x,\xi,{A}^{{(\lambda)}})
=A_\alpha^{{(\lambda)}\top} - \frac{\xi_\alpha^\top}{x\cdot\xi}(x\cdot A^{(\lambda)}).\;\;
\end{eqnarray}

\textbf{\emph{Part II.}} As already pointed out (see Eq. (\ref{xi xi =0}) and the associated discussion), the arbitrariness introduced with the spinors $u_1(\xi)$ and $u_2(\xi)$ can be removed by comparison with the flat case. In fact, following the lines sketched in Ref. \cite{Diracfields}, we have fixed them to get the usual solutions of the Dirac equation in Minkowski spacetime at the limit $H\rightarrow 0$. The Minkowskian limit of the spinors are \cite{Diracfields}
\begin{eqnarray}\label{posenspinor}
\lim_{H\rightarrow 0} u_{1,r}(\xi)=u_{1,r}(\vec{k})=\frac{\slashed{k}\gamma^4 +mc}{\sqrt{2mc(k^0+mc)}}u_r(\xi^\circ_+),
\end{eqnarray}
\begin{eqnarray}\label{negenspinor}
\lim_{H\rightarrow 0} u_{2,r}(\xi)=u_{2,r}(\vec{k})=\frac{-\slashed{k}\gamma^4 +mc}{\sqrt{2mc(k^0+mc)}}u_r(\xi^\circ_-),
\end{eqnarray}
where $\slashed{k}=k_\mu\gamma^\mu$, with $\mu=0,1,2,3$. Here, $\xi^\circ_\pm=(1,\vec{0},\pm 1)$ and
\begin{eqnarray}
u_r(\xi^\circ_+) =\frac{1}{\sqrt{2}} \left( \begin{array}{ccc}
\chi_r \\
\chi_r \end{array}\right), \;\;\;
u_r(\xi^\circ_-) =\frac{1}{\sqrt{2}} \left( \begin{array}{ccc}
\chi_r \\
-\chi_r \end{array} \right),\nonumber
\end{eqnarray}
with
\begin{eqnarray}
\chi_1 = \left( \begin{array}{ccc}
1 \\
0 \end{array}\right), \;\;\;
\chi_2 = \left( \begin{array}{ccc}
0 \\
1 \end{array} \right).\;\;\nonumber
\end{eqnarray}

\textbf{\emph{Part III.}} Finally, by using the parametrization (\ref{paraxi}) and (\ref{4.3}), the flat limit of the dS plane waves (\ref{coindSpw}), with homogeneity degree $\sigma_+$, is \cite{BrosPRL}
\begin{eqnarray}\label{posen}
\lim_{H\rightarrow 0}(Hx\cdot\xi_-)^{\sigma_+} &=& \exp (- ik\cdot X),\nonumber\\
\lim_{H\rightarrow 0} e^{-i\pi\sigma_+} (Hx\cdot\xi_+)^{\sigma_+} &=& \exp ( ik\cdot X),
\end{eqnarray}
where $\exp (\mp ik\cdot X)$ are the Minkowskian plane waves, respectively, with positive and negative energy: the contraction is indeed performed regarding the Lorentz subgroup $SO_0(1,3)$ ($\Gamma_4$ is invariant under $SO_0(1,3)$), hence the above equations demonstrate that the orbital basis $\Gamma_4$ contracts to the sum of two Minkowskian plane waves with opposite signs of energy \cite{Mickelsson}.

Here, we also must underline that, contrary to the Minkowskian exponentials, the dS plane waves, as functions on ${\cal M}_H$, are merely locally defined because they are singular on three-dimensional light-like manifolds, and moreover, since $(Hx\cdot\xi)$ can be negative,\footnote{Note that $x\cdot\xi_- >0$ and $x\cdot\xi_+ <0$.} they are multi-valued in dS spacetime. To get rid of these difficulties and obtain single-valued global waves, they have to be considered as distributions, as proposed in Refs. \cite{BrosPRL,Bros,BrosComm}, which are boundary values of analytic continuations of the dS plane waves to tubular domains in the complexified dS spacetime ${\cal M}^{(c)}_H$ defined by
\begin{eqnarray}
{\cal M}^{(c)}_H &=& \{ z=x+iy \in \mathbb{C}^5: \;\;\eta_{\alpha\beta}z^\alpha z^\beta = -H^{-2} \},\nonumber
\end{eqnarray}
where $\mathbb{C}^5$ is the ambient complex Minkowski spacetime.

There are distinguished domains of ${\cal M}^{(c)}_H$, namely, ${\cal T}^{\pm} = \mbox{T}^{\pm}\cap{\cal M}^{(c)}_H$ in which of them the global waves can be defined as analytic functions for $z$. Here $\mbox{T}^\pm= \mathbb{R}^5 \pm iV^+$ are, respectively, the forward and backward tubes in $\mathbb{C}^5$. Accordingly, the ``tuboid" is defined above ${\cal M}_H \times{\cal M}_H$ in ${\cal M}^{(c)}_H \times{\cal M}^{(c)}_H$ by
\begin{eqnarray}\label{tube}
{\cal T}^{\pm} = \{(z,z^\prime):\; z\in{\cal T}^{+}, z^\prime\in{\cal T}^{-}\}.
\end{eqnarray}

Let $\xi\in {\cal C}^+$ and $z\in{\cal T}^{+}$ (or ${\cal T}^{-}$), then for a generic $\sigma\in\mathbb{C}$, the plane waves
\begin{eqnarray}\label{x dot Xi}
\phi (z) = (Hz\cdot\xi)^\sigma,
\end{eqnarray}
would be defined globally, since in this case the imaginary part of $z\cdot\xi$ has a fixed sign and $z\cdot\xi \neq 0$. Now, the dS plane waves, with homogeneity degree $\sigma_+$, are given by \cite{GHR}
\begin{eqnarray}\label{bvofac}
\phi_+ (x) &\equiv&  e_\nu \mbox{\bf{bv}}(Hz\cdot\xi)^{\sigma_+} \nonumber\\
&=& e_\nu [\vartheta (Hx\cdot\xi) + \vartheta (-Hx\cdot\xi)e^{-i\pi{\sigma_+}}]\nonumber\\
&& \times |Hx\cdot\xi |^{\sigma_+},
\end{eqnarray}
where $\mbox{\bf{bv}}$ stands for the boundary value and $\vartheta$ is the Heaviside function. The real valued constant $e_\nu$ is specified by considering the Hadamard requirement on the corresponding two-point function. On this basis, for $\xi\in \Gamma_4$, we have \cite{GHR}
\begin{eqnarray}\label{limbv}
\lim_{H\to 0} \phi_+ (x) = \frac{1}{\sqrt{2(2\pi)^3}}e^{- ikX},
\end{eqnarray}
which means that the mode $\phi_+ (x)$ does not generate negative frequency mode in the limit $H\rightarrow 0$: whether $x\cdot\xi$ be positive or negative, we obtain (\ref{limbv}). It is worthy to point out that the above result, associating the positive energy with $\phi_+ (x)$, is obtained in the neighborhood of a fixed point $x$ in ${\cal M}_H$. Although, $\phi_+ (x)$ is globally defined in dS space, the concept of energy is not. In fact, under Bogoliubov transformations, one may obtain the conjugate modes $\phi_+^\ast$ whose flat limit at some point $x^\prime$ in ${\cal M}_H$ is a negative frequency mode when $x^\prime\cdot\xi <0$ \cite{GHR}.

Considering the above discussion, we define the four independent solutions (\ref{general solu 1}) and (\ref{general solu 2}) as the boundary value of the analytic continuation to
\begin{eqnarray}\label{comgeneral solu 1}
\Big({\psi^{(\frac{3}{2})}_{1,r}}\Big)_\alpha(z) = f{\cal E}^{(\lambda)}_{\alpha}(z,\xi) (Hz\cdot\xi)^{\sigma_-} u_{1,r}(\xi),\;\;\;\;\;
\end{eqnarray}
\begin{eqnarray} \label{comgeneral solu 2}
\Big({\psi^{(\frac{3}{2})}_{2,r}}\Big)_\alpha(z) = f^\ast{\cal E}^{(\lambda)}_{\alpha}(z,\xi) (Hz\cdot\xi)^{\sigma_+} \slashed{z} u_{2,r}(\xi),\;\;
\end{eqnarray}
where $\xi\in{\cal C}^+$ and $z\in{\cal T}^+$ (or ${\cal T}^{-}$) as in Eq. (\ref{x dot Xi}). Correspondingly, the solutions to the adjoint equation can be considered as the boundary value of the analytic continuation to
\begin{eqnarray} \label{comadj general solu 1}
\Big({\overline{\psi}^{\;(\frac{3}{2})}_{1,r}}\Big)_\alpha(z) = f^\ast{{\cal E}}^{\ast{(\lambda)}}_{\alpha}(z^\ast,\xi)(Hz\cdot\xi)^{\sigma_+} \overline{u}_{1,r}(\xi),\;\;\;
\end{eqnarray}
\begin{eqnarray} \label{comadj general solu 2}
\Big({\overline{\psi}^{\;(\frac{3}{2})}_{2,r}}\Big)_\alpha(z) = f{{\cal E}}^{\ast{(\lambda)}}_{\alpha}(z^\ast,\xi) (Hz\cdot\xi)^{\sigma_-} \overline{u}_{2,r}(\xi) \overline{\slashed{z}}.\;\;\;
\end{eqnarray}

In the next section, this family of the dS spinor-vector waves will be considered to define the Euclidean vacuum in the conventional terminology.

\subsection{Comment on massless limit}
We end our discussion in this section by commenting on the ``massless" limit of this construction which is much more elaborated. Recall that, the dS ``massless" spin-$\frac{3}{2}$ corresponds to the discrete series representations $\Pi_{\frac{3}{2},\frac{3}{2}}^\pm$, labeled by $\Delta=({p=\frac{3}{2},q=\frac{3}{2}})$. Therefore, to get the massless limit of this construction, we need to set $\nu=-i$ in the formulas presented above for which we have $N=-1$. As a direct consequence, the differential projection operator presented in (\ref{general solu}), on the classical level, becomes singular. This singularity is indeed because of the divergencelessness requirement that already imposed to relate the spinor-vector field $\psi^{(\frac{3}{2})}$ to a specific UIR of the dS group (see (\ref{suu})). To circumvent this difficulty, the divergencelessness requirement must be dropped. The field equation then becomes gauge invariant, namely $\psi^{(\frac{3}{2})}\rightarrow \psi^{(\frac{3}{2})} + D_1\psi^{(\frac{1}{2})}_g $ is a solution to the massless Rarita-Schwinger field equation for any spinor field $\psi^{(\frac{1}{2})}_g$, satisfying $\kappa\psi^{(\frac{1}{2})}_g = 0$, as far as $\psi^{(\frac{3}{2})}$ is. Accordingly, the general solutions transform under the dS indecomposable representations instead of the dS UIRs. Through the gauge-fixing procedure, the massless spin-$\frac{3}{2}$ field can ultimately be quantized. [In this regard, one may refer to a study of the anti-dS counterpart of the massless Rarita-Schwinger field given by Lesimple \cite{Lesimplemassless}.] We will discuss this matter in a forthcoming paper.

\section{Two-point function and quantum field}
In this section, following the concrete quantization procedure already presented and originally discussed for the scalar fields in the previous works by Bros et al \cite{Bros,BrosComm,BrosPRL}, we proceed to the quantization of the massive spinor-vector field in dS spacetime.

Let us begin our discussion by making the standard realization of a spinor-vector quantum field in dS space explicit. A spinor-vector quantum field $\Psi^{(\frac{3}{2})}(x)$ is, roughly speaking, expected to be an operator-valued distributions on ${\cal M}_H$ acting on (a suitable dense domain $D$ of) a Hilbert space ${\cal H}$. Besides positive definiteness, this definition of course should satisfy some physically reasonable properties. The quantum field should fulfill the locality requirement,
\begin{eqnarray}
\{\Psi^{(\frac{3}{2})}(x),\Psi^{(\frac{3}{2})}(x^\prime)\}&=&0,\nonumber\\
\{\Psi^{(\frac{3}{2})}(x),\overline{\Psi}^{\;(\frac{3}{2})}(x^\prime)\}&=&0,
\end{eqnarray}
for all space-like separated pair $(x,x^\prime)$ on ${\cal M}_H$ (i.e., $x\cdot x^\prime >-H^{-2}$). Here, as above, $\overline{\Psi}^{\;(\frac{3}{2})} = {\Psi^{(\frac{3}{2})}}^\dagger\gamma^0\gamma^4$. There also should exist a unitary representation ${U}(\Lambda_g)$ of dS group in ${\cal H}$ such that
\begin{eqnarray}
{U}(\Lambda_g)\; \Psi^{(\frac{3}{2})}(x) \;{U}(\Lambda_g)^{-1} &=& g^{-1} \Psi^{(\frac{3}{2})}(\Lambda_gx),\nonumber\\
{U}(\Lambda_g)\; \overline{\Psi}^{\;(\frac{3}{2})}(x)\; {U}(\Lambda_g)^{-1} &=& \overline{\Psi}^{\;(\frac{3}{2})}(\Lambda_gx)i(g),\nonumber
\end{eqnarray}
which means that $\Psi^{(\frac{3}{2})}$ and $\overline{\Psi}^{\;(\frac{3}{2})}$ transform covariantly. On the other hand, there should exist a normalized fundamental state $|{0}\rangle\in D$, called the vacuum, which is cyclic for the polynomial algebra of field operators and invariant under the dS representation, ${U}(\Lambda_g)|{0}\rangle = |{0}\rangle$. Moreover, the field should satisfy geodesical spectral condition or geometrical KMS condition which means that the vacuum is defined as a physical state with the temperature $T=\hbar cH/2\pi K$ ($K$ is the Boltzmann constant). Ultimately, the field should verify the conditions of transversality, divergencelessness, and tracelessness.

In this study, we are interested in the free field part of the theory for which all the truncated correlation functions vanish. The full information about the QFT is therefore entirely encoded in the two-point function,
\begin{eqnarray}
{S}^{(\frac{3}{2})}_{\alpha\alpha^\prime}(x,x^\prime) = \langle{0}|\Psi_\alpha^{(\frac{3}{2})}(x)\;\overline{\Psi}_{\alpha^\prime}^{\;(\frac{3}{2})}(x^\prime)|{0}\rangle,
\end{eqnarray}
with $\alpha,\alpha^\prime = 0,1,2,3,4$, which should be a spinor-vector valued distribution on ${\cal M}_H\times{\cal M}_H$ satisfying the following conditions:
\begin{itemize}
\item{\textbf{\emph{Positiveness.}} For any test function $h_\alpha \in {\cal D}(D{\cal M}_H)\sim {\cal D}({\cal M}_H)\bigotimes \mathbb{C}^4$, the positiveness condition necessitates the inequalities
\begin{eqnarray}\label{Positiveness}
\langle h,h\rangle = \int_{{\cal M}_H\times {\cal M}_H} d\Omega(x)d\Omega(x^\prime)\hspace{1.7cm} \nonumber\\
\times {{\overline{h}}}^{\;\alpha} (x) {S}^{(\frac{3}{2})}_{\alpha\alpha^\prime}(x,x^\prime) h^{\alpha^\prime}(x^\prime)\geq 0,
\end{eqnarray}
in which $d\Omega(x)$ stands for the dS-invariant measure on ${\cal M}_H$.}
\item{\textbf{\emph{Covariance.}} For any $g\in Sp(2,2)$, the two-point function is dS invariant, i.e.,
\begin{eqnarray}
g^{-1}{S}^{(\frac{3}{2})}_{\alpha\alpha^\prime}(\Lambda_gx, \Lambda_gx^\prime)i(g) &=& {S}^{(\frac{3}{2})}_{\alpha\alpha^\prime}(x, x^\prime).
\end{eqnarray}}
\item{\textbf{\emph{Local anticommutativity.}} For all space-like separated events $(x,x^\prime)$ on ${\cal M}_H$, the locality requirement implies that
\begin{eqnarray}
{S}^{(\frac{3}{2})}_{\alpha\alpha^\prime}(x,x^\prime) &=& \langle{0}|\Psi_\alpha^{(\frac{3}{2})}(x)\;\overline{\Psi}_{\alpha^\prime}^{\;(\frac{3}{2})}(x^\prime)|{0}\rangle \nonumber\\
&=& - \langle{0}|\overline{\Psi}_{\alpha^\prime}^{\;(\frac{3}{2})} (x^\prime)\; \Psi_\alpha^{(\frac{3}{2})}(x)|{0}\rangle\nonumber\\
&=& - \widehat{S}^{(\frac{3}{2})}_{\alpha^\prime\alpha}(x^\prime,x),
\end{eqnarray}
where $\widehat{S}^{(\frac{3}{2})}_{\alpha^\prime\alpha}(x^\prime,x) = \langle{0}|\overline{\Psi}_{\alpha^\prime}^{\;(\frac{3}{2})} (x^\prime)\; \Psi_\alpha^{(\frac{3}{2})}(x)|{0}\rangle$ is the ``permuted" two-point function.}
\item{\textbf{\emph{Transversality.}}
\begin{eqnarray}
x^\alpha{S}^{(\frac{3}{2})}_{\alpha\alpha^\prime}(x, x^\prime) = 0 = {S}^{(\frac{3}{2})}_{\alpha\alpha^\prime}(x, x^\prime){x^\prime}^{\alpha^\prime}.
\end{eqnarray}}
\item{\textbf{\emph{Divergencelessness.}}
\begin{eqnarray}
\partial_x^\alpha{S}^{(\frac{3}{2})}_{\alpha\alpha^\prime}(x, x^\prime) = 0 = {S}^{(\frac{3}{2})}_{\alpha\alpha^\prime}(x, x^\prime)\overleftarrow{\partial}_{x^\prime}^{\alpha^\prime}.
\end{eqnarray}
From now on, for the sake of simplicity, we define $\partial^\prime \equiv\partial_{x^\prime}$.}
\item{\textbf{\emph{Tracelessness.}}
\begin{eqnarray}
\gamma^\alpha{S}^{(\frac{3}{2})}_{\alpha\alpha^\prime}(x, x^\prime) = 0 = {S}^{(\frac{3}{2})}_{\alpha\alpha^\prime}(x, x^\prime)(-\gamma^4 {\gamma}^{\alpha^\prime}\gamma^4).
\end{eqnarray}}
\item{\textbf{\emph{Normal analyticity.}} The two-point function ${S}^{(\frac{3}{2})}_{\alpha\alpha^\prime}(x, x^\prime)$ is the boundary value (in the distributional sense) of an analytic function ${S}^{(\frac{3}{2})}_{\alpha\alpha^\prime}(z, z^\prime)$ in the domain ${\cal T}^{\pm}$ (see (\ref{tube})).}
\end{itemize}

From the normal analyticity condition, stated above, we have: (\emph{i}) ${S}^{(\frac{3}{2})}_{\alpha\alpha^\prime}(z, z^\prime)$ is maximally analytic, which means that it can be analytically continued to the ``cut domain"
\begin{eqnarray}
\Delta =\{(z,z^\prime)\in {\cal M}_H^{(c)}\times {\cal M}_H^{(c)}\; :\; (z-z^\prime)^2 < 0 \};
\end{eqnarray}
(\emph{ii}) The permuted function $\widehat{S}^{(\frac{3}{2})}_{\alpha^\prime\alpha}(x^\prime,x)$ is the boundary value of $\widehat{S}^{(\frac{3}{2})}_{\alpha^\prime\alpha}(z^\prime,z)$ in the domain ${\cal T}^{\mp} = \{(z,z^\prime);\; z\in{\cal T}^{-}, z^\prime\in{\cal T}^{+}\}$ of ${\cal M}_H^{(c)}\times {\cal M}_H^{(c)}$.\footnote{Note that the permuted two-point function satisfies all the above listed requirements, as well.}

Once the above requirements are met, the reconstruction theorem \cite{Streater} allows one to construct a QFT satisfying all the axioms. On this basis, our present task would be to find such a doubled spinor-vector valued analytic function of the variable $(z,z^\prime)$ exhibiting the above properties.

To achieve this goal, considering the framework presented thus far, we formally expand the field operators $\Psi^{(\frac{3}{2})}(z)$ and $\overline{\Psi}^{\;(\frac{3}{2})}(z)$, for any given value of mass parameter $\nu$, as follows
\begin{eqnarray}\label{11111}
\Psi^{(\frac{3}{2})}(z) &=& ff^\ast\int_\Gamma d\Omega_\Gamma (\xi) \sum_{r}\sum_{\lambda} \nonumber\\
&&\Big\{a_r(\xi) {{\cal E}^{(\lambda)}}(z,\xi)(Hz\cdot\xi)^{\sigma_-} u_{r}(\xi) \nonumber\\
&&+ d_r^\dagger(\xi) {{\cal E}^{(\lambda)}}(z,\xi)(Hz\cdot\xi)^{\sigma_+} \slashed{z}u_{r}(\xi) \Big\},\hspace{0.7cm}
\end{eqnarray}
\begin{eqnarray}\label{22222}
\overline{\Psi}^{\;(\frac{3}{2})}(z) &=& ff^\ast\int_\Gamma d\Omega_\Gamma (\xi) \sum_{r}\sum_{\lambda} \nonumber\\
&& \Big\{a^\dagger_r(\xi){\cal E}^{\ast (\lambda)}(z^\ast,\xi) (Hz\cdot\xi)^{\sigma_+} \overline{u}_{r}(\xi)\nonumber\\
&&+ d_r(\xi) {\cal E}^{\ast (\lambda)}(z^\ast,\xi) (Hz\cdot\xi)^{\sigma_-} \overline{u}_{r}(\xi)\overline{\slashed{z}} \Big\},\hspace{0.7cm}
\end{eqnarray}
where $a$, $a^\dagger$, $d$ and $d^\dagger$ are operator-valued amplitudes, and $d\Omega_\Gamma (\xi)$ stands for the natural ${\cal C}^+$ invariant measure on $\Gamma$, induced from the $\mathbb{R}^5$ Lebesgue measure \cite{Bros}. The operators $a$ and $d$, defining by $a_r(\xi)|0\rangle = d_r(\xi)|0\rangle = 0$, satisfy the canonical anticommutation relations.

For any positive $\nu$ (the principal-series parameter), the above expansions allow us to write the analytic two-point function ${S}^{(\frac{3}{2})}(z,z^\prime)$ explicitly in terms of the following class of integral representation
\begin{eqnarray}\label{int 2-point}
{S}^{(\frac{3}{2})}(z,z^\prime) &=& ff^\ast c_\nu \int_\Gamma d\Omega_\Gamma(\xi) (Hz\cdot\xi)^{\sigma_-} (Hz^\prime\cdot\xi)^{\sigma_+}\nonumber\\
&\times& \sum_{r} u_{r}(\xi)\overline{u}_{r}(\xi) \sum_{\lambda} {\cal E}^{(\lambda)}(z,\xi) {\cal E}^{\ast (\lambda)}({z^\prime}^\ast,\xi)\nonumber\\
&=& ff^\ast c_\nu \int_\Gamma d\Omega_\Gamma(\xi) (Hz\cdot\xi)^{\sigma_-} (Hz^\prime\cdot\xi)^{\sigma_+}\nonumber\\
&\times & \Big(\frac{1}{2}\slashed{\xi}\gamma^4\Big) \sum_{\lambda} {\cal E}^{(\lambda)}(z,\xi) {\cal E}^{\ast (\lambda)}({z^\prime}^\ast,\xi),
\end{eqnarray}
which is valid for all points in the tuboid ${\cal T}^\pm$. Here, $c_\nu$ is a normalization constant which is specified by imposing the Hadamard condition on the two-point function. The Hadamard condition indeed determines a unique vacuum state for the spinor-vector quantum field verifying the field equation.

In order to check whether Eq. (\ref{int 2-point}) fulfills the aforementioned requirements, it would also be convenient to rewrite the two-point function in a more suitable form. On this basis, by substituting Eq. (\ref{E 1'}) into (\ref{int 2-point}), and with respect to the identities given in previous section (see (\ref{3.17})), we develop the two-point function (\ref{int 2-point}) and find
\begin{eqnarray}\label{2point-xi}
{S}_{\alpha\alpha^\prime}^{(\frac{3}{2})}(z,z^\prime) &=& -ff^\ast c_\nu \int_\Gamma d\Omega_\Gamma(\xi)\nonumber\\
&&\times \Big[(\theta_{\alpha}\cdot\theta^\prime_{\alpha^\prime})
- H^{-2} {\cal Z} \frac{\xi^\top_\alpha \xi^\top_{\alpha^\prime}}{(z\cdot\xi)(z^\prime\cdot\xi)} \nonumber\\
&& - \Big( (\theta_{\alpha}\cdot {z^\prime}) \frac{\xi^\top_{\alpha^\prime}}{(z^\prime\cdot\xi)}+ (z\cdot \theta^\prime_{\alpha^\prime})\frac{\xi^\top_{\alpha}}{(z\cdot\xi)} \Big)\Big]\nonumber\\
&& \times\Big(\frac{1}{2}\slashed{\xi}\gamma^4\Big) (Hz\cdot\xi)^{\sigma_-} (Hz^\prime\cdot\xi)^{\sigma_+}.
\end{eqnarray}
in which ${\cal Z} = - H^2 z\cdot z^\prime$ is a dS-invariant variable.

Considering the following relation
\begin{eqnarray}
H^2 D_1 (Hz\cdot\xi)^{\sigma_\pm} = \frac{{\sigma_\pm}\xi^\top(Hz\cdot\xi)^{\sigma_\pm}}{(z\cdot\xi)},\nonumber
\end{eqnarray}
the two-point function (\ref{2point-xi}) can be expressed as the resulting action of a differential operator on the dS Dirac-spinor analytic two-point function ${S}^{(\frac{1}{2})}(z,z^\prime)$ as
\begin{eqnarray}\label{2-point D}
{S}^{(\frac{3}{2})}_{\alpha\alpha^\prime}(z,z^\prime) = {\mathscr E}_{\alpha\alpha^\prime} (z,z^\prime) {S}^{(\frac{1}{2})}(z,z^\prime),
\end{eqnarray}
with
\begin{eqnarray}
{\mathscr E}_{\alpha\alpha^\prime} (z,z^\prime) &=& -ff^\ast \Big[ (\theta_\alpha\cdot\theta^\prime_{\alpha^\prime}) - {H^{2}}\Big( \frac{{\cal Z} D_{1\alpha}D^\prime_{1\alpha^\prime}}{\sigma_-\sigma_+}\Big)\nonumber\\
&& - H^2 \Big( \frac{(z\cdot\theta^\prime_{\alpha^\prime})D_{1\alpha}}{\sigma_-} + \frac{(z^\prime\cdot\theta_{\alpha})D^\prime_{1\alpha^\prime}}{\sigma_+} \Big)\Big],\nonumber
\end{eqnarray}
and
\begin{eqnarray}
{S}^{(\frac{1}{2})}(z,z^\prime) &=& \langle{0}|\Psi^{(\frac{1}{2})}(z) \; \overline{\Psi}^{\;(\frac{1}{2})}(z^\prime) |{0}\rangle \nonumber\\
&=& e^2_\nu \int_\Gamma d\Omega_\Gamma(\xi) \Big(\sum_{r} u_r(\xi)\overline{u}_r(\xi)\Big)\nonumber\\
&&\hspace{1.5cm}\times(Hz\cdot\xi)^{\sigma_-} (Hz^\prime\cdot\xi)^{\sigma_+} \nonumber\\
&=& e^2_\nu \int_\Gamma d\Omega_\Gamma(\xi) \Big( \frac{1}{2}\slashed{\xi}\gamma^4 \Big) \nonumber\\
&&\hspace{1.5cm}\times(Hz\cdot\xi)^{\sigma_-}(Hz^\prime\cdot\xi)^{\sigma_+}.\;\;\;\;\;
\end{eqnarray}
Note that: (\emph{i}) The primed operators act only on the primed coordinates and vise versa; (\emph{ii}) The above differential operator ${\mathscr E}(z,z^\prime)$ clearly verifies
\begin{eqnarray}\label{locality identity}
{\mathscr E}^\ast(z^\ast,{z^\prime}^\ast) ={\mathscr E}(z,z^\prime).
\end{eqnarray}
This property will serve to show the local anticommutativity requirement; (\emph{iii}) For well chosen space-like separated points $z$ and $z^\prime$ in the domain ${\cal T}^\pm$, the explicit form of the ${S}^{(\frac{1}{2})}(z,z^\prime)$ is proportional to a generalized Legendre function \cite{Diracfields},
\begin{eqnarray} \label{spinor 2-point}
{S}^{(\frac{1}{2})}(z,z^\prime) = -\frac{1}{8}E_\nu\Big[{\sigma_-}P^{(7)}_{\sigma_+}(-{\cal Z})\slashed{z}\hspace{2cm}\nn\\
- {\sigma_+} P^{(7)}_{\sigma_-}(-{\cal Z})\slashed{z}^\prime\Big]\gamma^4,
\end{eqnarray}
where $E_\nu = 2i\pi^2 e^{\pi\nu} e^2_\nu$ and
\begin{eqnarray}
e^2_\nu = \frac{\nu(1+\nu^2)}{(2\pi)^3(e^{2\pi\nu} - 1)}.\nn
\end{eqnarray}

Eventually, taking the boundary value of ${S}^{(\frac{3}{2})}(z,z^\prime)$ from ${\cal T}^{\pm}$, one can express the integral (Fourier type) representation of the two-point function, in terms of the dS global waves on the real hyperboloid ${\cal M}_H$, as follows
\begin{eqnarray} \label{bvspinor 2-point}
{S}^{(\frac{3}{2})}(x,x^\prime) &=& \mbox{\bf{bv}} {S}^{(\frac{3}{2})}(z,z^\prime) \nonumber\\
&=&ff^\ast c_\nu \sum_\lambda \int_\Gamma d\Omega_\Gamma(\xi) \nonumber\\
&& \times \mbox{\bf{bv}} \Big((Hz\cdot\xi)^{\sigma_-} (Hz^\prime\cdot\xi)^{\sigma_+}\Big)\nonumber\\
&& \times {\cal E}^{(\lambda)}(x,\xi) \Big(\frac{1}{2}\slashed{\xi}\gamma^4\Big){\cal E}^{\ast (\lambda)}({x^\prime},\xi),\;\;\;
\end{eqnarray}
where, according to (\ref{bvofac}),
\begin{eqnarray}
\mbox{\bf{bv}} \Big((Hz\cdot\xi)^{\sigma_-} (Hz^\prime\cdot\xi)^{\sigma_+}\Big)=\hspace{4cm} \nn\\
|Hx\cdot\xi |^{\sigma_-}|Hx^\prime\cdot\xi |^{\sigma_+} [\vartheta (Hx\cdot\xi) + \vartheta (-Hx\cdot\xi)e^{i\pi{\sigma_-}}] \nn\\
\times [\vartheta (Hx^\prime\cdot\xi) + \vartheta (-Hx^\prime\cdot\xi)e^{-i\pi{\sigma_+}}].\nn
\end{eqnarray}

Now, considering the boundary value limit, it can be proved that the presented kernel in (\ref{bvspinor 2-point}) satisfies the aforementioned conditions required in order to get a Wightman two-point function. The existence of the latter is indeed requested by dS axiomatic field theory.
\begin{itemize}
\item{In order to show the positiveness property, by making use of the Fourier-Bros transformation on ${\cal M}_H$ \cite{BrosComm}, we consider the Fourier transform on ${\cal M}_H$ of each spinorial component $h_i(x)$ of $h(x)$,
    \begin{eqnarray}
    \widetilde{h}_i^{(\lambda)}(\xi) = f^\ast\int_{{\cal M}_H} d\Omega(x) {\cal E}^{\ast (\lambda)}_{\alpha}({x},\xi) h^{\alpha}_i(x)\hspace{.5cm}\nonumber\\
    \times [ \vartheta(Hx\cdot\xi) + \vartheta(-Hx\cdot\xi)e^{+i\pi {\sigma_-}} ] |Hx\cdot\xi|^{\sigma_-}.
    \end{eqnarray}
    Accordingly, the integral given in the positiveness requirement, see (\ref{Positiveness}), takes the following form
    \begin{eqnarray}
    \langle h,h \rangle = c_\nu \int_{\Gamma} d\Omega_\Gamma(\xi) \sum_{\lambda} \overline{\widetilde{h}}^{(\lambda)}(\xi) \Big( \frac{1}{2}\slashed{\xi}\gamma^4 \Big) \widetilde{h}^{(\lambda)}(\xi).
    \end{eqnarray}
    Noting that $c_\nu$ is positive, and for all $\xi$ on the integration manifold the matrix $\gamma^0\gamma^4\slashed{\xi}\gamma^4$ is positive semi-definite \cite{Diracfields}, the positiveness requirement is met, i.e., $\langle h,h \rangle \geq 0$.}
\item{In order to prove the local anticommutativity requirement, we need to identify the anticommutator for space-like separated points. Following the identity (\ref{locality identity}) and considering the permuted two-point function corresponding to the massive dS Dirac field given by \cite{Diracfields}
    \begin{eqnarray}
    \widehat{S}^{(\frac{1}{2})}(z^\prime,z) = -\slashed{z}{S}^{(\frac{1}{2})}(z^\prime,z) \gamma^4 \slashed{z}^\prime \gamma^4,\nonumber
    \end{eqnarray}
    which in combination with (\ref{spinor 2-point}) yields $\widehat{S}^{(\frac{1}{2})}(z^\prime,z) = -{S}^{(\frac{1}{2})}(z,z^\prime)$, one can easily see that, for space-like separated points, the anticommutator operator vanishes,
    \begin{eqnarray}
    {S}^{(\frac{3}{2})}(z,z^\prime) + \widehat{S}^{(\frac{3}{2})}(z^\prime,z)=0.\nonumber
    \end{eqnarray}
    This fulfills the locality requirement. [Recall that the space-like separated pair $(x,x^\prime)$ lies in the same orbit of the complex dS group as the pairs $(z,z^\prime)$ and $({z^\prime}^\ast,z^\ast)$.]}
\item{The covariant transformations of the dS spinor-vector modes (\ref{covariance}) and the independence of the integral (\ref{int 2-point}) from the chosen orbital basis $\Gamma$ and from the corresponding measure $d\Omega_\Gamma(\xi)$ explicitly guarantee the covariance property.}
\item{The analyticity properties of the two-point function realize from the expression of the spinor-vector plane waves (\ref{comgeneral solu 1}) - (\ref{comadj general solu 2}).}
\item{The transversality, divergencelessness, and tracelessness with respect to $x$ and $x^\prime$ is assured because the dS spinor-vector modes $\psi^{(\frac{3}{2})}(x)$ and the corresponding adjoint ones $\overline{\psi}^{\;(\frac{3}{2})}(x)$ are transverse, divergenceless, and traceless by construction.}
\end{itemize}

Here, it is worth noting that the normal analyticity requirement that has been considered above will play the role of a spectral condition in the absence of a global energy-momentum interpretation in dS spacetime. Indeed, this requirement, along the lines proposed in \cite{Bros,BrosComm,BrosPRL} and \cite{Diracfields}, implies a thermal-KMS interpretation.

Another remarkable advantage of this construction that should be pointed out here is that the above formulas allow a factorization of the two-point function $S^{(\frac{3}{2})}(x,x^\prime)$ in terms of the dS global plane waves which, in the null-curvature limit, is explicitly analogous to the associated Fourier representation for the two-point function of the Minkowski Rarita-Schwinger free field with the Poincar$\mbox{\'{e}}$ mass $m$. Technically, regarding the orbital basis $\Gamma_4$, the Minkowskian limit is straightforward to calculate: the measure $d\Omega_{\Gamma_4}(\xi)$ is considered to be $m^2$ times the natural one induced from the $\mathbb{R}^5$ Lebesgue measure, i.e., $d\Omega_{\Gamma_4}(\xi)=d^3\vec{k}/k^0$. On this basis, we have \cite{Diracfields}
\begin{eqnarray}
\lim_{H\rightarrow 0} S^{(\frac{1}{2})}(x,x^\prime) &\simeq& \int \frac{d^3\vec{k}}{(2\pi)^3} \frac{m}{k^0} e^{-ik(x-x^\prime)} \frac{\slashed{k}\gamma^4 + m}{2m}\;\nonumber\\
&\equiv& S_{M}^{(\frac{1}{2})}(x,x^\prime),\nonumber
\end{eqnarray}
and
\begin{eqnarray}
\lim_{H\rightarrow 0} \widehat{S}^{(\frac{1}{2})}(x^\prime,x) &\simeq& - \int \frac{d^3\vec{k}}{(2\pi)^3} \frac{m}{k^0} e^{ik(x-x^\prime)} \frac{\slashed{k}\gamma^4 - m}{2m}\nonumber\\
&\equiv& \widehat{S}_{M}^{(\frac{1}{2})}(x^\prime,x),\nonumber
\end{eqnarray}
where the subscript $M$ refers to the Minkowskian counterpart. Then, considering Eq. (\ref{2-point D}) along the identities given in the previous sections, the Minkowskian limit of $S^{(\frac{3}{2})}(x,x^\prime)$ and $\widehat{S}^{(\frac{3}{2})}(x^\prime,x)$, respectively, reads
\begin{eqnarray}
\lim_{H\rightarrow 0} \left \{ \begin{array}{rl} S^{(\frac{3}{2})}(x,x^\prime)\vspace{2mm}\\
\vspace{2mm} \widehat{S}^{(\frac{3}{2})}(x^\prime,x)\end{array}\right \} = \hspace{5cm}\nonumber\\
- \Big( \eta_{\mu\nu} + \frac{1}{m^2} \frac{\partial^2}{\partial X^\mu \partial X^\nu} \Big) \left \{ \begin{array}{rl} S_M^{(\frac{1}{2})}(x,x^\prime)\vspace{2mm}\\
\vspace{2mm} \widehat{S}_M^{(\frac{1}{2})}(x^\prime,x)\end{array}\right \}.\nonumber
\end{eqnarray}

At the end, let us close this section by making the associated Hilbert space structure explicit. Indeed, the explicit knowledge of the two-point function ${S}^{(\frac{3}{2})}(x,x^\prime)$ fulfilling the above-mentioned requirements, with respect to the reconstruction theorem \cite{Streater}, allows us to justify the introduction of the dS massive spinor-vector field $\Psi^{(\frac{3}{2})}$, that is, an operator-valued distributions on ${\cal M}_H$ verifying the dS Rarita-Schwinger field equation (\ref{Field Eq. newwwwww}) and acting on (a dense domain of) a separable Hilbert space ${\cal H}$. The latter, with positive-definite metric, can be described as the direct sum
\begin{eqnarray}
{\cal H} = {\cal H}_0 \oplus \Big[ \oplus_{n=1}^\infty {\cal A} {\cal H}_1^{\otimes n} \Big],
\end{eqnarray}
in which ${\cal A}$ is the antisymmetrization operator and
\begin{eqnarray}
{\cal H}_0 = \{ \digamma |{0}\rangle,\; \digamma \in \mathbb{C} \}.\nonumber
\end{eqnarray}
With respect to creation and annihilation operators, the field operators in smeared form $\Psi^{(\frac{3}{2})}(g)$, on the (dense) class of regular elements $h\in {\cal H}_1$ and for each test function $h_\alpha(x) \in {\cal D}(D{\cal M}_H)$, reads
\begin{widetext}
\begin{eqnarray}
\Big(\Psi^{(\frac{3}{2})}(g) h\Big)^{(n)} (x_1,\alpha_1,i_1;\; x_2,\alpha_2,i_2; \;.\;.\;.\;;x_n,\alpha_n,i_n) = \hspace{8cm}\nonumber\\
\frac{1}{\sqrt{n}} \sum_{k=1}^{n} (-1)^{k+1} g_{\alpha_k,i_k}(x_k) h^{(n-1)} (x_1,\alpha_1,i_1;\;.\;.\;.\;;\breve{x}_k,\breve{\alpha}_k,\breve{i}_k; \;.\;.\;.\; ;x_n,\alpha_n,i_n)\nonumber\\
+ \sqrt{n+1} \int_{{\cal M}_H \times {\cal M}_H} d\Omega(x)d\Omega(x^\prime) g_{\alpha,i}(x){\Big(S^{(\frac{3}{2})}}^{\alpha\alpha^\prime,i\;\overline{i^\prime}}(x,x^\prime)\Big) h^{(n+1)} (x^\prime,\alpha^\prime,\overline{i^\prime};\;x_1,\alpha_1,i_1;\;.\;.\;.\;;x_n,\alpha_n,i_n).
\end{eqnarray}
\end{widetext}
Note that: (\emph{i}) $\breve{x}_k,\breve{\alpha}_k,\breve{i}_k$ means that these terms are omitted; (\emph{ii}) $i$ and $\overline{i^\prime}$ stand for spinorial indices.

\section{Conclusion}
In this paper, we have quantized the ``massive" Rarita-Schwinger field in dS spacetime by adapting to this specific situation the content of previous works: the ambient space notations, construction of the dS plane waves (the modes), and finally construction of the Wightman two-point function which leads to the covariant quantization of the theory.

Our group theoretical description of the dS massive Rarita-Schwinger field, verifying the Wightman axioms and supplemented by analyticity requirements in the complexified pseudo-Riemannian manifold, once again confirms that the Euclidean vacuum has to be preferred vacuum in dS spacetime as far as one wishes to recover the usual QFT in the null-curvature limit. More precisely, it shows that the Euclidean vacuum of the dS massive Rarita-Schwinger field is the only vacuum for which the Minkowskian limit of the theory leads to, \emph{at any point of spacetime}, positive frequency modes. [Of course, this does not mean that the energy concept is defined globally in dS spacetime. Indeed, any Bogoliubov transformation on the given modes at the point $x$ may result in the appearance of modes at some point $x^\prime$ with negative frequency in the flat limit.] Moreover, the use of the dS plane waves in our quantization approach reveals that the whole free dS massive Rarita-Schwinger field theory tends toward the corresponding QFT in Minkowski spacetime when the curvature vanishes, including the dS Fourier transform which becomes the standard one in the limit. On the other hand, it is also worth noting that the analyticity properties of the spinor-vector plane waves and the two-point function, which have been introduced in this paper, allow for a detailed study of the Hilbert space of the massive Rarita-Schwinger field in dS spacetime, and along the lines sketched in Refs. \cite{Bros,BrosComm,BrosPRL} and \cite{Diracfields}, give rise to the thermal physical interpretation of the theory.

As a final remark, concerning the link with the involved UIR's, we point out that the Rarita-Schwinger equation (\ref{Field Eq. newwwwww}), resulting from the eigenvalue equation (\ref{eigenfieq}), acts on fields which carry a finite direct sum of UIR's (including $U_{\frac{3}{2},\nu}$) and not only $U_{\frac{3}{2},\nu}$. Indeed, we require both, the quadratic and the quartic, dS Casimir operators to specify in general the involved representations. [Recall that, in this paper, we only consider the quadratic one.] In this sense, it would be useful to examine the action of the quartic dS Casimir operator on the field solutions. This feature is in relation with the existence of redundant components. These extra components are responsible for the acausal propagation of solutions in Minkowski spacetime in the presence of interactions, a pathology put in evidence by the famous papers \cite{Velo01,Velo02} (see also Refs. \cite{Hagen,Hurley,Wightman3/2,Gazeau3/2}). Understanding whether the above pathologies can be cured by some extent in dS spacetime, when the action of both dS Casimir operators are taken into account, would be an interesting extension of the present work.

\section*{Acknowledgements}
This work was supported by the National Natural Science Foundation of China with the Grants Nos: 11375153 and 11675145.

\begin{appendix}

\setcounter{equation}{0}
\section{\label{factorization}Spacetime factorization of the group}
To be able to understand the structure of spacetime, one may factorize $g$ into two parts $g=jl$. This procedure would be based on the involution $g\rightarrow i(g) = \gamma^0\gamma^4 g^\dagger \gamma^0\gamma^4$, and we call it spacetime factorization.

Here, the factor $l$ is an element of the (Lorentz) subgroup,
\begin{eqnarray}
L=\{l\in Sp(2,2) :\; l\;i(l)=\mathbb{I}\}\simeq SL(2,\mathbb{C}).\nonumber
\end{eqnarray}
This factor leaves the origin of dS spacetime $O_H$ invariant, i.e.,
\begin{eqnarray}
\left( \begin{array}{ccc}
0 & 1\\
1 & 0 \end{array} \right) = l\left( \begin{array}{ccc}
0 & 1\\
1 & 0 \end{array} \right)l^\dagger, \nonumber
\end{eqnarray}
and takes the form
\begin{eqnarray}
l=\left( \begin{array}{ccc}
\zeta & 0\\
0 & \zeta \end{array} \right) \left( \begin{array}{ccc}
\cosh (\varphi_1 /2) & {w}\sinh (\varphi_1 /2)\\
-{w}\sinh (\varphi_1 /2) & \cosh (\varphi_1 /2) \end{array} \right), \nonumber
\end{eqnarray}
where $\varphi_1 \in \mathbb{R}$, $\zeta\in SU(2)$ and the parameter ${w}\in SU(2)$ is the ``pure" vector quaternion ($w=-w^\star$). In this sense, the parameters $\zeta$, ${w}$ and $\varphi_1$ are respectively presumed to carry the meaning of space rotation, boost velocity direction, and rapidity.

The factor $j$, on the other hand, maps the origin to any point of dS spacetime, i.e.,\footnote{We recall the following correspondence between points of the hyperboloid ${\cal M}_H$ and $2\times 2$ quaternionic matrices
\begin{eqnarray}
{\cal M}_H \ni x \rightarrow \slashed{x}\equiv x\cdot\gamma = \left( \begin{array}{ccc}
x^0 & {\cal P}\\
{\cal P}^\star & x^0 \end{array} \right)\gamma^0 =\left( \begin{array}{ccc}
x^0 & -{\cal P}\\
{\cal P}^\star & -x^0 \end{array} \right),\nn
\end{eqnarray}
with ${\cal P}\equiv (x^4,\vec{x})=x^4+x^ke_k\in\mathbb{H}$ and ${\cal P}^\star=(x^4,-\vec{x})$.}
\begin{eqnarray}
j\left( \begin{array}{ccc}
0 & 1\\
1 & 0 \end{array} \right)j^\dagger = \left( \begin{array}{ccc}
x^0 & {\cal P}\\
{\cal P}^\star & x^0 \end{array} \right). \nonumber
\end{eqnarray}
The factor $j$ can be explicitly characterized by
\begin{eqnarray}
j=\left( \begin{array}{ccc}
\tau & 0\\
0 & {\tau}^\star \end{array} \right) \left( \begin{array}{ccc}
\cosh (\varphi_2 /2) & \sinh (\varphi_2 /2)\\
\sinh (\varphi_2 /2) & \cosh (\varphi_2 /2) \end{array} \right), \nonumber
\end{eqnarray}
with $\varphi_2\in\mathbb{R}$ and $\tau\in SU(2)$, for which, we have
\begin{eqnarray}
j\;i(j) &=& \left( \begin{array}{ccc}
\tau^2\cosh \varphi_2 & \sinh\varphi_2\\
\sinh\varphi_2 & {(\tau^2)}^\star\cosh\varphi_2 \end{array} \right)\nn\\
&\equiv &
\left( \begin{array}{ccc}
x^0 & -{\cal P}\\
{\cal P}^\star & -x^0 \end{array} \right)
\left( \begin{array}{ccc}
0 & 1\\
-1 & 0 \end{array} \right)=\slashed{x}\gamma^4.
\nn
\end{eqnarray}
This equivalence holds modulo a determinant factor, which means that $j$ is a kind of ``spacetime" square root, which exemplifies the dS topology $S^3\times \mathbb{R}$: the set $\{ \varphi_2, \tau^2 \}$ provides global coordinates for ${\cal M}_H$ through
\begin{eqnarray}
x^0=\sinh\varphi_2, \;\;\;\; \mbox{and} \;\;\;\; {\cal P}= \tau^2\cosh\varphi_2.\nn
\end{eqnarray}

\section{\label{realization}$S^3$ realization}
The UIRs of the principal series on the hyperboloid ${\cal M}_H$ can be constructed from their realizations on a spherical section $\Gamma_0$ of the cone ${\cal C}$ ($S^3$ realization). The Hilbert space carrying these representations are the spaces $L^2(SU(2))$. The action of $g = \left( \begin{array}{ccc}
a & b\\
c & d \end{array} \right)\in Sp(2,2)$ on $\tau \in SU(2)$ is given by $\tau^\prime = g\cdot\tau = (a\tau + b)(c\tau + d)^{-1}$. Let $\Phi_0\in L^2(SU(2))$, then the action of $Sp(2,2)$ on $\Phi_0$ can be represented in the following form
\begin{eqnarray}
(U_{s,\nu}(g)\Phi_0)(\tau) = |c\tau + d|^{-2\ell} D^s \Big(\frac{{\tau}^\star {c}^\star+{d}^\star}{|c\tau + d|}\Big)\Phi_0(g^{-1}\cdot\tau),\nn
\end{eqnarray}
with $g^{-1}= \left( \begin{array}{ccc}
a & b\\
c & d \end{array} \right)\in Sp(2,2)$ and $\ell = (3/2) + i\nu $. $D^s$ is the UIR of $SU(2)$ of dimension $(2s+1)$, $\tau = (\xi^4, \vec{\xi})$, $\tau\cdot\tau = 1$ and $\xi \in \Gamma_0$. For a transformation which maps this realization to the realization with the hyperbolical orbital basis $\Gamma_4$ used in this paper, we refer the reader to \cite{Taka}.

\end{appendix}


\begin{thebibliography}{a}

\bibitem{Linde} A. Linde, \emph{Particle Physics and Inflationary Cosmology}, Harwood Academic Publishers, Chur (1990).

\bibitem{Perlmutter} S. Perlmutter et al, Astro. Phys. J. 483, 565 (1997); B. Schmidt et al, Astro. Phys. J. 507, 46 (1998); A. J. Riess et al, Astron. J. 116, 1009 (1998).

\bibitem{Maldacena} J. Maldacena, Adv. Theo. Math. Phys. 2, 231 (1998) [hep-th/9711200]; E. Witten, Adv. Theor. Math. Phys. 2, 253 (1998); S. Gubser, I. Klebanov and A. Polyakov, Phys. Lett. B 428, 105 (1998); O. Aharony, S. Gubser, J. Maldacena, H. Ooguri and Y. Oz, Phys. Rep. 323, 183 (2000) [hep-th/9905111].

\bibitem{Witten} E. Witten, \emph{Quantum gravity in de Sitter space}, hep-th/0106109.

\bibitem{Strominger} A. Strominger, JHEP 0110, 034 (2001).

\bibitem{Hull} C.M. Hull, JHEP 9807, 021 (1998) [hep-th/9806146]; Mu-In Park, Phys. Lett. B 440, 275 (1998); Nucl. Phys. B 544, 377 (1999); I. Antoniadis, P. Mazur and E. Mottola, astro-ph/9705200; A. Volovich, hep-th/0101176; V. Balasubramanian, P. Horava and D. Minic, JHEP, 0105, 043 (2001).

\bibitem{BrosComm} J. Bros, H. Epstein, and U. Moschella, Commun. Math. Phys. 196, 535 (1998).

\bibitem{Bros} J. Bros, U. Moschella, Rev. Math. Phys. 08, 327 (1996).

\bibitem{BrosPRL} J. Bros, J.P. Gazeau, U. Moschella, Phys. Rev. Lett. 73, 1746 (1994).

\bibitem{Chernikov1968} N.A. Chernikov, and E. A. Tagirov. Ann. lnst. Henri Poincar\'{e} A9, 109 (1968).

\bibitem{Tagirov73} E. A. Tagirov, Ann. Phys. (N.Y.) 76, 561 (1973).

\bibitem{Schom68} J. G\'{e}h\'{e}niau and C. Schomblond, Acad. R. Belg. Bull. Cl. Sci. 54, 1147 (1968).

\bibitem{Schom76} C. Schomblond and P. Spindel, Ann. lnst. Henri Poincar\'{e} A25, 67 (1976).

\bibitem{Mottola84} E. Mottola, Phys. Rev. D 31, 754 (1984).

\bibitem{Allen85} B. Allen, Phys. Rev. D 32, 3136 (1985).

\bibitem{de Bievre6230} S. De Bi\`{e}vre and J. Renaud, Phys. Rev. D 57, 6230 (1998).

\bibitem{Gazeau1415} J.P. Gazeau, J. Renaud, and M.V. Takook, Class. Quant. Grav. 17, 1415 (2000).

\bibitem{masive/vect} J.P. Gazeau and M.V. Takook, J. Math. Phys. 41, 5920 (2000).

\bibitem{Diracfields} P. Bartesaghi, J.P. Gazeau, U. Moschella, and M.V. Takook, Class. Quant. Grav. 18, 4373 (2001).

\bibitem{massive/2} T. Garidi, J.P. Gazeau, and M.V. Takook, J. Math. Phys. 44, 3838 (2003).

\bibitem{Garidi/2005} T. Garidi, E. Huguet, and J. Renaud, J. Phys. A 38, 245 (2005).

\bibitem{massless/vect} T. Garidi, J.P. Gazeau, S. Rouhani, and M.V. Takook, J. Math. Phys. 49, 032501 (2008).

\bibitem{dSGravityII} H. Pejhan and S. Rahbardehghan, Phys. Rev. D 94, 104030 (2016).

\bibitem{dSGravityI} H. Pejhan and S. Rahbardehghan, Phys. Rev. D 93, 044016 (2016).

\bibitem{BambaI} K. Bamba, S. Rahbardehghan and H. Pejhan, Phys. Rev. D 96, 106009 (2017).

\bibitem{massless/2} H. Pejhan, K. Bamba, S. Rahbardehghan, and M. Enayati, Phys. Rev. D 98, 045007 (2018).

\bibitem{massless/22} H. Pejhan, M. Enayati, J.P. Gazeau, and A. Wang, Phys. Rev. D 100, 066012 (2019).

\bibitem{Mickelsson} J. Mickelsson and J. Niederle, Commun. Math. Phys. 27, 167 (1972).

\bibitem{Dooley} A.H. Dooley, \emph{Contractions of Lie groups and applications to analysis, in: ``Topics in modern harmonic analysis"}, vol.1, 483-515 (Instituto Nationale di Alta Matematica Francesco Serveri), Roma (1983).

\bibitem{GHR} T. Garidi, E. Huguet and J. Renaud, Phys. Rev. D 67, 124028 (2003).

\bibitem{barut} A.O. Barut , A. B\"ohm, J. Math. Phys. 11, 2938 (1970).

\bibitem{What} T. Garidi, \emph{What is mass in desitterian physics?}. arXiv preprint hep-th/0309104 (2003).

\bibitem{AFFS} E. Angelopoulos, M. Flato, C. Fronsdal, and D. Sternheimer, Phys. Rev. D 23, 1278 (1981).

\bibitem{WG} A.S. Wightman, and L. Garding, \emph{Fields as operator-valued distributions in relativistic quantum theory}, Arkiv Fys. 28 (1965).

\bibitem{Unruh} W.G. Unruh, Phys. Rev. D 14, 870 (1976).

\bibitem{Sewell} G.L. Sewell, Ann. Phys. 141, 201 (1982).

\bibitem{Hawking} S.W. Hawking, Commun. Math. Phys. 43, 199 (1975).

\bibitem{Basu} A. Basu, L.I. Uruchurtu, Class. Quant. Grav. 23, 6059 (2006).

\bibitem{Flato1978} M. Flato, and C. Fronsdal, Lett. Math. Phys. 2, 421 (1978).

\bibitem{Hidden} H. Pejhan, S. Rahbardehghan, M. Enayati, K. Bamba, and A. Wang, Phys. Lett. B 795, 220 (2019).

\bibitem{Grensing} G. Grensing, J. Phys. A: Math. Gen. 10, 1687 (1977).

\bibitem{Moylan} P. Moylan, J. Math. Phys. 24, 2706 (1983).

\bibitem{Dix} J. Dixmier, Bull. Soc. Math. Fr. 89, 9 (1961).

\bibitem{Taka} B. Takahashi, Bull. Soc. Math. Fr. 91, 289 (1963).

\bibitem{Lipsman} R.L. Lipsman, \emph{Group representations: a survey of some current topics}, Vol. 388, Springer (1974).

\bibitem{Mizony} M. Mizony, Publ. Dep. Math. Lyon. 3, 47 (1984).

\bibitem{FaFr} J. Fang and C. Fronsdal, Phys. Rev D 22, 1361 (1980).

\bibitem{Gazeau88} J.P. Gazeau and M. Hans, J. Math. Phys. 29, 2533 (1988).

\bibitem{Lesimple} M. Lesimple, Lett. Math. Phys. 15, 143 (1988).

\bibitem{GelV} I.M. Gel'fand, and G. Vilenkin, Academic Press (1964) \emph{Generalized Functions}, Vol. V.

\bibitem{Lesimplemassless} M. Lesimple, Lett. Math. Phys.  18, 315 (1989).

\bibitem{Streater} R.F. Streater, A.S. Wightman, and W.A. Benjamin, Inc. (1964) PCT, \emph{Spin and Statistics, and All That}.

\bibitem{Velo01} G. Velo, and D. Zwanziger, Phys. Rev. 186, 1337 (1969).

\bibitem{Velo02} G. Velo, and D. Zwanziger, Phys. Rev. 188, 2218 (1969).

\bibitem{Hagen} C.R. Hagen, and W.J. Hurley, Phys. Rev. Lett. 24, 1381 (1970).

\bibitem{Hurley} W.J. Hurley, Phys. Rev. Lett. 29, 1475 (1972).

\bibitem{Wightman3/2} A.S. Wightman, \emph{Studies in Mathematical Physics, Essays in Honor of Valentine Bargmann}, Princeton University Press p. 423, (1976).

\bibitem{Gazeau3/2} J.P. Gazeau, J. Phys. G. 6, 1459 (1980); 7, 1311 (1981).

\end{thebibliography}
\end{document}